\documentclass[12pt,preprint]{aastex}
\usepackage{natbib}
\usepackage{amsmath}
\begin{document}

\title{Constraining A String Gauge Field by Galaxy Rotation Curves and  Perihelion Precession  of Planets}
\author{Yeuk-Kwan~E.~Cheung\altaffilmark{2} and Feng Xu\altaffilmark{1}}
\affil{Department of Physics, Nanjing University, 22 Hankou Road, Nanjing, China 210093}
\altaffiltext{1}{Current address: Department of Physics, Universit\"{a}t Bonn, D-53115 Bonn, Germany}
\altaffiltext{2}{Correspondance: [cheung@nju.edu.cn]}

\begin{abstract}
We discuss a cosmological model in which the string gauge field coupled universally to matter gives rise to an extra centripetal force and will have observable signatures  on cosmological and astronomical observations. Several tests are performed using data including galaxy rotation curves of twenty-two spiral galaxies of varied luminosities and sizes, and perihelion precessions of planets in the solar system.
The rotation curves of the same group of galaxies are independently fit using a dark matter model with the generalized Navarro--Frenk--White (NFW) profile and  the string model.
Remarkable fit of galaxy rotation curves is achieved using the one-parameter string model as compared to the three-parameter  dark matter model with the Navarro-Frenk-White profile.  
The average $\chi^2$ value of the NFW fit  is 9\% better than that of the string model at a price of two more free parameters. 
Furthermore, from the string model, we can  give a dynamical explanation for  the phenomenological Tully-Fisher relation. 
We are able to derive a relation between field strength, galaxy size and luminosity, which can be verified with data from the 22 galaxies.
To further test the hypothesis of the universal existence of the string gauge field, 
we apply our string model to 
the solar system. 
Constraint on the magnitude of the string  field in the solar system  is deduced from the current  ranges for any anomalous perihelion precession of planets  allowed by the latest observations. 
The field distribution resembles a dipole field originating from the Sun.  
The string field strength deduced from the solar system observations 
is of a similar magnitudes as the field strength needed to sustain the rotational speed of the sun inside the Milky Way. This hypothesis can be  tested further  by future observations with higher precision. 


\end{abstract}

\maketitle

\section{Introduction}
\addcontentsline{toc}{section}{Introduction}

\label{sec:introduction}

Recent cosmological  and astronomical observations  are becoming increasingly interesting laboratories for precision tests of new physical theories aimed at extending the standard paradigms. On the one hand, the high-energy completion of gravity theory should leave signatures sufficiently different from the low-energy effective theories  similar to  general relativity and its modifications.  These  will be detectable with the advances in detector technologies. 
On the other hand, string theory and other quantum gravity  candidate theories  when applied to cosmology or astronomy,  should  shed new light on old problems. Therefore it is extremely important and timely to work out the observable signatures from various quantum gravity theories, as summarized in a recent review~\citep{Hossenfelder:2010zj}.

In this work, we will confine our interest to  two problems.
The first one  is the ``missing mass problem'' in galaxies--the  discrepancy  between mass  measured by rotational speeds of stars inside a spiral galaxy and mass predicted from its stellar matter distribution. 
Another problem  is the recently reported anomalous precession of  planets inside the solar system, the  explanation of which cannot be found within the standard framework.
The model we propose to solve these two problems at the same time  is a very special string model. The model,  first discovered by   
Nappi \& Witten~\citep{Nappi:1993ie}, 
falls into a general class of exactly solvable string models but it has the added merit that all effects due to the finite size of the strings are taken into account.  In other words it is  a {\it bona fide} string model, and it lives in four dimension spacetime which  closely resembling  Minkowski spacetime but with the presence of the  string gauge field.   
Due to the existence of the string gauge field, the geodesics  in this 4D space-time are concentric circles instead of the usual  straight lines in Minkowski spacetime. 
This is analogous to the Laudau orbits of an electron in the presence of a magnetic field. Thus Cheung, Kao and Savvidy~\citep{CheungGRC0:2007} proposed the model to explain the galaxy rotation curves in spiral galaxies in lieu of Cold Dark Matter (CDM). 
In this paper we  extend their work 
with a direct comparison of the goodness of fit of the string model with the CDM model in galaxy rotation curves fitting.  
Furthermore we  apply the idea to the solar system planetary perihelion precessions in order to infer further constraints on the model's parameters.

\section{Galaxy rotation curve}
\label{sec:GRC}
While the CDM cosmology has been accepted by many as the correct theory for structure formation on a  large scale and the solution to the missing mass problem on the galactic scale we still lack hard proof for the existence of Dark Matter particles. 
Until the coming Linear Hadron Collider and future experiments--see exciting development in this direction~\citep{Chang:2008zzr,Adriani:2008zr}--tells us definitely what constitute Dark Matter, a more natural and universal explanation in lieu of dark matter cannot be excluded. 
Here we entertain the possibility that a higher-rank gauge field universally coupled to strings can give rise to  a Lorentz force in four dimensions providing an extra centripetal acceleration for matter towards the center of a galaxy in addition to the gravitational attraction due to stellar matter.
If not properly accounted for, it would appear as if there were extra {\it{invisible}} matter in a galaxy. A salient feature of this Lorentz-like force is that it  fits  galaxies with an extended region of linearly rising rotational velocity significantly better than the dark matter model.  
This feature also endows the model with testability: in the region where the gravitational attraction of the visible matter completely gives way to the linear rising Lorentz force, typically in the region $ r\sim 20R_{d}$,  should we still observe  linear rising rotation velocity, say for the satellites of the host galaxy, it would be a strong support for the model. 
Otherwise if  {\it all} rotation curves  are found to  fall off  beyond $  20R_{d}$ for all galaxies then the model is proven wrong.  
Furthermore this is the first  such attempt to directly  verify the validity of string theory as a description of  low energy physics.   
Given this last  reason alone we regard it as a worthwhile endeavour.

\subsection{The string model}
\label{subsec:string_model}
Consider a four-dimensional string model proposed by Nappi and Witten~\citep{Nappi:1993ie} in which  the string theory is exactly integrable.   Furthermore   tree-level correlation functions--describing an arbitrary number of interacting particles--have been computed, which capture  all finite-sized effects of the strings to all orders in the string  scale~\citep{Cheung:2003ym}. 
This  is valid for all energy scales as long as the string coupling constant is weak.  This is the region of interest when we extrapolate to our low energy world.  The three-form background  gauge field,  $H_{(3)}$,  coupled uniquely to the worldsheet of the strings,  is constant.   

Because we are no longer approximating strings as point particles, this coupling between the two-form gauge potential  $B_{(2)}$ and the two dimensional worldsheet of the string produces a net force on the string when it is  viewed as a point particle~\citep{Cheung:2003ym}.
The center of mass of the closed string executes Landau orbits given by:
\begin{equation}  
\label{eq: landau}
a = a_0 + r  e^{i\Omega\,t}~,
\end{equation}
where $a$ is the complex coordinate of the plane in which the time-like part of the three-form field has non-zero components. 
This phenomenon is completely  analogous to the behaviour of an electron in  a constant magnetic field. When applied to model galaxy rotation curve, $a $ parameterizes the galactic plane  while  $H$ permeates the whole galaxy, and farther beyond,  but always crosses the plane parameterized by $a$ at right angle.

For the purpose of the data fitting in this paper 
we only need to retain a component of the tensor gauge field which is perpendicular to the galactic plane of the spiral galaxies~\footnote{%
In a refinement of the model we let this string gauge field be generated by the rotating stellar matter, and gases, itself. The profile of the string gravimagnetic field generated by  a rotating disk 
of stellar matters is the same as the magnetic field profile generated  by an electrically charged disk.}.
We denote the strength of this gauge field by $H$. Together with the ``charge-mass'' ratio, it forms the {{\it only}} free parameter of this model, denoted by    $\Omega \equiv \frac{Q}{m} H$.  
This is to be contrasted with the three free parameters one needs in the dark matter model using the celebrated 
Navarro--Frenk--White (NFW) profile~\citep{NFW} for an iso-thermal and isotropic dark matter distribution,   $r_{s}$ (the characteristic radius of the dark matter distribution), $\sigma_{0}$ (the central density), and $\alpha$ (the steepness parameter).   
A general profile with a free parameter $\alpha$ is used because of the need to fit rotation curves from dwarf galaxies as well as   galaxies with a varied surface brightness, from high surface brightness to low surface brightness. 

The simplistic  property of the string model  is, in fact, a favoured approach from  the string theory point of view, because the coupling of the field to matter has a universal strength, i.e. all matter is charged, rather than neutral. 
Therefore if all matter is indeed made up of fundamental strings and hence couples universally to  the tensor gauge field, $H$, each star in a  galaxy  will then execute  the circular motion in  concentric landau orbits on the galactic plane.  
Effectively there is an additional Lorentz force term in the equation of motion for a test star, the sum of the usual gravitational force and an additional Lorentz force term due to $H$:

\begin{eqnarray}  \label{eq:totalforce}
\vec{F}_{\mathrm{total}}&=&\vec{F}_{string}+\vec{F}_{\mathrm{gravity}}\\
&=&q\vec{v}\times\vec{H}+\vec{F}_{\mathrm{gravity}}
\end{eqnarray}
whose radial component reads
\begin{eqnarray}  \label{eq: EOM}
m\,  \frac{v^2}{r} = q \,H\,v + \,  m\, F_{stellar}~
\end{eqnarray}
where the field, $\vec{H}$, is generated by the rotating stellar matter and the halo of  gases alike and has a profile of a magnetic field generated by a rotating electrically-charged disk. One can easily verify  that $\vec{v}  \times \vec{H}$ contributes to an additional centripetal acceleration. 
$F_{stellar}$ denotes the contributions from visible, stellar matter of a spiral galaxy which will be explained in detail in section~\ref{subsec:stellar_matter}. The stellar contribution  is  in common  between the cold dark matter  model and the string model.

Let us pause to remark that  our model  is different from other series of models called ``celestial ephemerides'' aiming to replace the Minkowski background with FRW background with an isotropic Hubble flow where there is an additional isotropic radial velocity, 
see for example~\citep{Kopeikin:2012iy,Iorio:2012wva}. 
In particular, note that in their work the force is proportional to the \emph{radial} component of the orbital velocity, while the one used
in our work is proportional to the \emph{transversal} component. We should also  note  here that modified Newtonian dynamics (MOND); see, for example,~\citep{Famaey:2011kh} for a sample of original literatures and the latest developments in this direction) is another approach to explain the galaxy rotation curves in lieu of CDM. 

While we are not claiming that we are replacing the CDM paradigm with the string field alone, we are alerting the readers of the possibility  that the  string gauge field, 
under which all matter--electrically neutral or not--is charged, can account for, at least partially, the inexplicable rotational speed of stars around the center of a spiral galaxy. However in the work we are presenting here we are pushing the limit of our proposal: in the ``string'' model we are not allowing any dark matter component at all and use one free parameter to fit the same set of spiral galaxies which are  independently fit to the CDM model  with three free parameters; and we compare the goodness of fit.

\subsection{The Dark Matter Model}    
\label{subsec:dark_matter_model}
According to the CDM paradigm  there is approximately 10 times more dark matter than visible matter.  
The fluctuations of the primordial density perturbations of the universe get amplified by gravitational instabilities.  
Hierarchical clustering models  further   predict that dark matter density traces the density of the universe  at the time of collapse and thus all dark matter halos have similar densities.    
Baryons  then fall into the gravitational potential created by the clusters of  dark matter particles,  forming the visible part of the galaxies. 
In a galaxy the dark matter exists in a spherical halo engulfing all of the visible matter and  extending  much further beyond the stellar disk.  
To describe the dark matter component we use the generalized NFW profile:
\begin{equation}        
\label{eq: NFW}
\sigma =\frac{\sigma_0 } { (\frac{r}{r_s})^\alpha (1 + \frac{r}{r_s})^{3-\alpha}}~,
\end{equation}
where $\alpha=1$ corresponds to the NFW profile, and  
$r_s$ is the characteristic radius of the dark matter halo.  
In the dark matter fitting routine we allow  $r_s$ to vary from $3R_d$ to $30R_d$.  
We further require that the dark matter density  be strictly  smaller than the visible mass
density, 
 $\sigma_0 < \rho_0$.  (The data can in fact be fit equally well when  the roles of dark matter and
visible matter inverted.)
Here we treat $\sigma_0$, $r_s$ and $\alpha$ as free parameters.  
Together with $ \rho_0$ and $R_d$ from  the visible component,  the dark matter model 
utilizes five free parameters.
All in all the rotation velocity of a test star is given by 
\begin{equation}  \label{eq:DM}
\frac{v^2}{r} =   F_{stellar}  +  F_{DM} 
\end{equation}  
in the dark matter model.

\subsection{Stellar Matter}
\label{subsec:stellar_matter}
To describe the visible matter we use the parametric distribution with exponential fall off in density from   
van der Kruit and Searle in both models:
\begin{equation}        
\label{eq: vdKS}
\rho  (r, z) = \rho_0\, \displaystyle{ e^{-\, \frac{r}{R_d}}}\, sech^2(\frac{z}{Z_d})
\end{equation}
with $\rho_0$ being the central matter density, $R_d$  the characteristic radius of the stellar disc and $Z_d$ the characteristic thickness.  
Following a common practice we choose $Z_d$ to be $\frac{1}{6} R_d$; the dependence of the final results   on this choice is very weak~\citep{vandenBosch:2000rza}.

Gravitational attraction due to the visible matter is  henceforth given by  
\begin{equation} \label{eq:Fstar}
F_{stellar}({r})
 =G_{N}\, \rho_0\, R_{d}\,  \tilde{F} (\tilde{r}) 
\end{equation}
where after rescaling $\tilde{r} \equiv \frac{r}{R_{d}}$
$$
\tilde{F}(\tilde{r})= \frac{\partial }{\partial  \tilde{r}} \, \int_{all\, space}
\displaystyle{\frac{e^{-r'}sech^2(6z')}{|\vec{ \tilde{r}}-\vec{r'}|}}~r'dr'dz'd\phi' 
$$ 
becomes a  universal function  for all galaxies.

\paragraph{Summary of equations in both models:}
We are now ready to fit the galaxy rotation curves data.   In the string model the three free parameters, $\Omega$, $R_d$, and $\rho_0$  are defined by the following equation:
\begin{equation}                  
\label{eq:eom_string}
\frac{v^2}{r}  = G_{N} \rho_0\,  R_{d}\,  \tilde{F} + 2\, \Omega\,  v ~.  
\end{equation}  
The fundamental charge-to-mass ratio  and the strength of the gauge field is encoded altogether in {\em one } free parameter $\Omega \equiv \frac{qH}{2m}$.  

The five free parameters in CDM model are defined by 
\begin{equation}  \label{eq:eom_DM}
\frac{v^2}{r} =  G_{N} \rho_0\,  R_{d}\,  \tilde{F}  +  F_{DM}~.
\end{equation} 
The two free parameter $\rho_0$\, and   $R_{d}$  are common to
 both models, describing stellar contribution to the rotational velocities of stars about the center of galaxy.
$F_{DM}(r)$ is given by the following expression:
$$
F_{DM}(r) = 4\pi \int^{r} dr' \frac{\sigma_0 } { (\frac{r'}{r_s})^\alpha (1 + \frac{r'}{r_s})^{3-\alpha}}~.
$$
All in all the  CDM profile carries another three free parameters, namely,  the central density of dark matter halo, $\sigma_0$, the  characteristic length scale of the halo, $R_{s}$, as well as the ``steepness'' parameter of the halo, $\alpha$.

\subsection{Fitting Procedures:}
A few remarks concerning our fitting procedures are in order.
The Dark Matter model with its five free parameters  and the string model  with its three free parameters are {{\it independently}} fit to the  data to obtain the best fit values for each set of the parameters for each of the twenty-two spiral galaxies.  
Under {{\it no}} circumstances are the best fit values from one model  fed into the other model as prior values.  
Except restricting  the $r_s$ to vary from $3R_d$ to $30R_d$ in CDM model  and letting $Z_d$ to be $\frac{1}{6} R_d$ in the stellar distribution  as conventionally done (see for example~\citep{vandenBosch:2000rza}) to save computing time there are no other simplifications.
We then use the best fit values for these two parameters (and three others in the dark matter model) to compute the total mass,  as well as the mass-to-light ratios, for these galaxies.     
 These will serve as sanity check for the best fit values of the parameters in  both models.

Independent of any galaxy rotation modelling, $\rho$ and $R_d$ can be fit with  photometric data and hence are not really free parameters.   Since we are only interested in comparing  the dark matter model and our string model  in  fitting the galaxy rotation curves we are treating them as free parameters in each model. We instead choose to  use our best fit values for these parameters, from Dark Matter model and String model  in turn, to compute the total mass in each galaxy and cross check with  independent astronomical observations.  We also compute the percentages of baryonic matter in the galaxies  for the CDM model, which is commonly done by astronomers.  This serves as another check of our methodology.

We are ignoring the gas contributions from our fitting; because putting in more free parameters will no doubt  improve the fit for both models.   
For the same reason we do not allow for any correction for star extinctions and supernova feedback  as they  would not affect any conclusion we draw  concerning the {{\it relative}} quality of the fit between the CDM model and string model.   
Keeping this simplistic spirit  we do not allow for any dark matter component at all in the string  model  and we also  assume that the strength of the string field be constant throughout the span of each  galaxy.   
Local back reaction of spacetime  to the presence of the string field is also ignored.

\subsection{Analysis}
\label{subsec:analysis}
The rotation velocity of a given test star is solved from equations  
(\ref{eq:eom_string}) and (\ref{eq:eom_DM}) 
for the string and the dark matter model, respectively.
The data are then fit to the dark matter model and string model independently.  The best fit values
of the parameters from each model are 
 obtained by minimizing the $\chi^2$ functionals\footnote{Note that in the observations the distance measurement from the center of the galaxy is assumed to be exact. 
  The uncertainty is instead attributed to the velocity measurements. During fitting, however,  we discovered that uncertainty in determining  the ``center of the galaxy'' significantly affects the quality of the fit.
According to both models, the rotation velocity at the ``center'' of the galaxy should be exactly zero.  
If we could shift some data by a linear translation to make the zero velocity point coincide with the $r=0$ point by hand, we would have obtained much lower $\chi^2$ values for both models.  Therefore this linear shift  is better attributed to the error in distance determination.}.  %
We obtained our rotation curve  data of  the twenty-two galaxies in the SINGS sample from the FaNTOmM website.
Using the best fit values of these parameter of the dark matter model, we can compute the mass of the dark matter halo, the mass of stellar mass, and then the ratio of luminosity to the total mass.
These values are tabulated in  Table~\ref{fig-best_fit_DM} of Appendix~\ref{appendix_RC}. 
From the string model we can likewise determine the set of values for the strength of the string  gauge field, the stellar mass and then finally the ratio of luminosity to the stellar mass. 
These values are tabulated in  Table~\ref{fig-best_fit_string} in    Appendix~\ref{appendix_RC}.

In  Figure~\ref{fig:string_ngc2403_example}, the rotation curve of  galaxy NGC2403 fitted using the  NFW profile   (left) and 
the   string model (right) are  plotted side by side for comparison. 
Squares with error bars are observational data.  
Theoretical predictions are indicated by the solid lines with stars in the NFW fit (left) and with triangles in the string fit (right).  The string
model clearly gives a better fit.

\begin{figure}

\begin{center}
$
\begin{array}{cc}
\includegraphics[width=75mm, height=60mm]{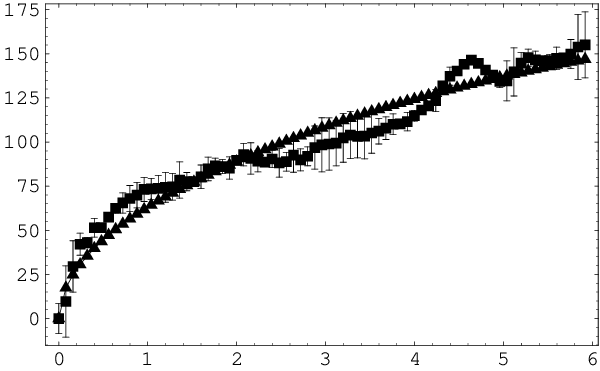}
&
\includegraphics[width=72mm, height=55mm]{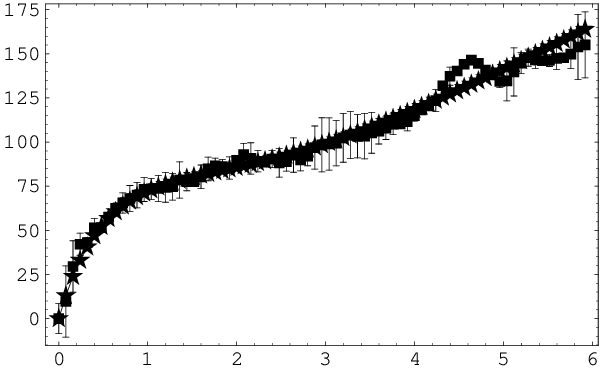}
\\
\end{array}
$
\end{center}
\caption{\label{fig:string_ngc2403_example} Rotation curve of NGC2403 fit with the dark matter model   (left) and with the  string model  (right).  
The $\chi$-squared value per degree of freedom using the dark matter model with a NFW profile is $4.515$ while that using  the string model is $4.304$.  
The X-axis is radius in $kpc$ and Y-axis is velocity  in $kms^{-1}$.  
}
\end{figure}  

The $\chi^{2}$ value, per degree of freedom, from  the string  fit is $4.304$ whereas that  from the NFW  fit is $4.515$.
Overall  string model gives a $\chi^{2}$ value  of  $1.656$ averaged over the 22 galaxies while the dark matter model gives a value of $1.594$. The fitting results  of 22 galaxies using the  dark matter model and the string model are detailed in Appendix~\ref{appendix_RC}.  The best fit values of the free parameters are tabulated in Table~\ref{fig-best_fit_DM} for the dark matter  model and 
in Table~\ref{fig-best_fit_string} for the string model, respectively, in Appendix~\ref{appendix_RC}.     We can see that the NFW profile  fits marginally better  at a  price of   two more  free parameters.

After we obtain the best fit values for the  free parameters we can compute the (total) masses for the galaxies. 
 
\paragraph{String Model:}  For this model there is only visible matter whose mass can be straightforwardly computed by integrating (\ref{eq: vdKS}) with the best fit values of $R_{d}$ and $\rho_{0}$ for each galaxy.
\paragraph{NFW profile:}

Matter in this model consists of the visible matter, same as that in the string model,  and the dark matter which  assumes the generalized NFW density profile~(\ref{eq: NFW}). 
The NFW profile gives divergent mass if the radius is integrated to infinity.  We therefore adopt the usual cutoff and compute the mass only up to the virial radius within which the average density is 200 times the critical density for closure.

\subsubsection{Visible Mass-to-Light versus $B$-Magnitude}
\begin{figure}[h!]
\begin{center}
\includegraphics[width=88mm, height=70mm]{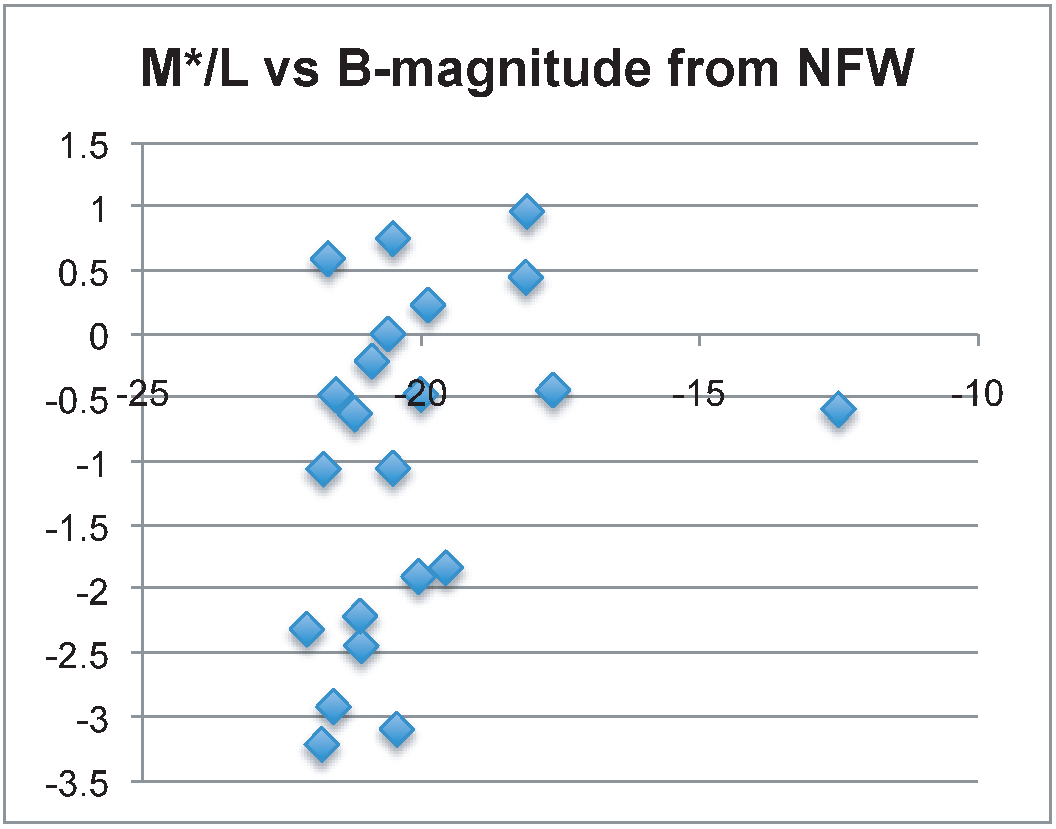}  
\end{center}
\vspace{-0.5cm}
\caption{\label{fig:DM_mass-to-light} The total-mass-to-light ratios derived from the best fit values and the measured B-band
luminosity of the 22 galaxies  in dark matter model. 
}
\end{figure}

\begin{figure}[h!]
\begin{center}
\vspace{0.5cm}
\includegraphics[width=88mm, height=70mm]{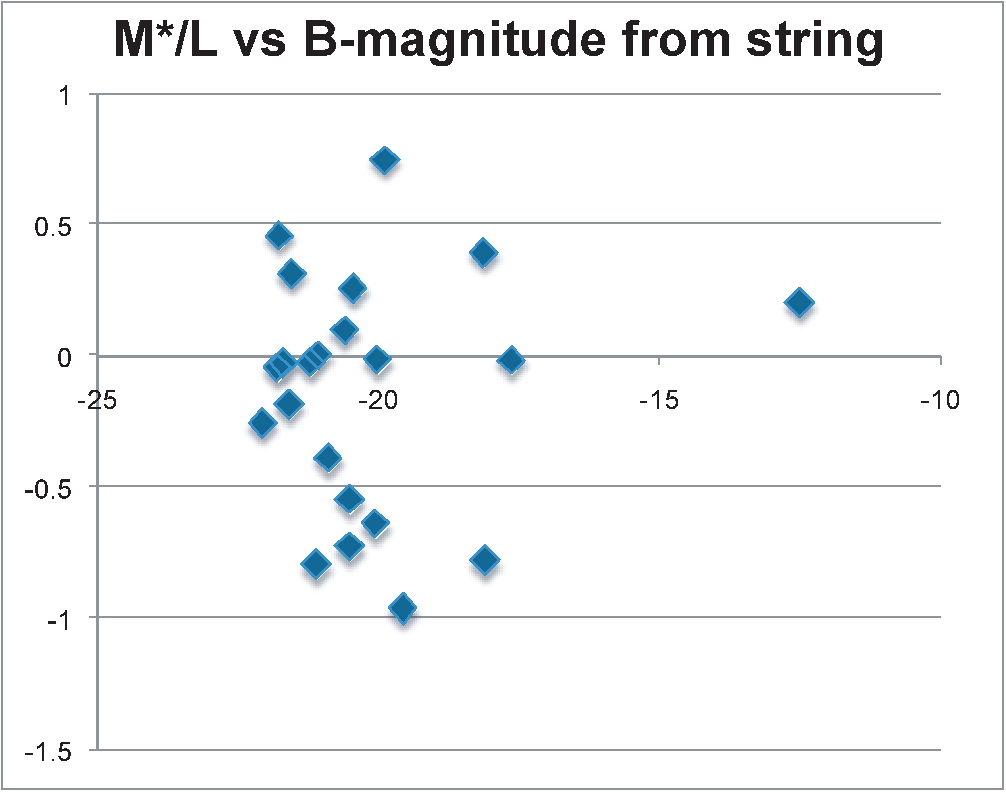} 
\end{center}
\vspace{-0.5cm}
\caption{\label{fig:string_mass-to-light} The visible-mass-to-light ratios derived from the best fit values and the measured B-band luminosity of the 22 galaxies in the string model. 
}
\end{figure}

Using the measured $B$-band absolute magnitudes we  compute the visible mass to light ratios for the
galaxies.  In the string model these ratios  fall between 
0.11 and  5.6  centered  around $1$ as shown
in Figure~\ref{fig:string_mass-to-light}. 
 The same ratios from the NFW model span five  orders of magnitude (see Figure~\ref{fig:DM_mass-to-light}) with many of them falling far below $1$.  
For the NFW model we also compute the percentage  of  baryonic matter in the total mass.  According to the CDM paradigm this number should be around $10\%$.  However the actual results are quite scattered.  The  scatter in the mass-to-light ratios and the baryon fractions clearly indicate that the NFW profile is not capturing the underlying physics correctly.  

\subsubsection{The Tully-Fisher Relation}
A Tully-Fisher relation  can be derived from the string model which relates the rotation velocity in
the ``flat'' region of the rotation curves to the product of the total luminous mass, $M_{stellar}$, and
the parameter, $\Omega\,$,
\begin{equation}
\label{eq: TF}
v^{3} = GM_{stellar} \Omega~.
\end{equation}
From the equation of motion~(\ref{eq: EOM}) we  solve for $v$,
\begin{equation} \label{eq: v}
v = \Omega\, r\, + \, \sqrt{ \Omega^{2}\, r^{2}\, + F_{stellar} \, r }~.
\end{equation}

We then look for a balance of falling Newtonian attraction and rising Lorentz force, resulting in 
$\frac{ \partial v} {\partial r} \sim 0$.  Because we know that the turning point is at $r \sim 2.2
r_{d}$,  setting $\frac{ \partial v} {\partial r} \sim 0$ yields  a relation between  $r_{d}$ and 
$\Omega$:
\begin{equation} 
\label{eq: omega}
8 \, \Omega^{2}\, \sim \frac{GM_{stellar}}{r^{3}}~.
\end{equation}
Inside the orbit  $r \sim 2.2R_{d}$ lies  most of  the visible mass. We can therefore use the
point-mass approximation when computing $F_{stellar}$ and $\frac{ \partial F_{stellar}} {\partial r}$.  
Upon substituting~(\ref{eq: omega}) into~(\ref{eq: v})   our Tully-Fisher relation  follows.  The
string model therefore provides a dynamic origin of this well-tested rule of thumb.
\begin{figure}[h!]  
\label{fig:Tully-Fisher}
\vspace{0.5cm}
\begin{center}
\includegraphics[width=80mm, height=70mm]{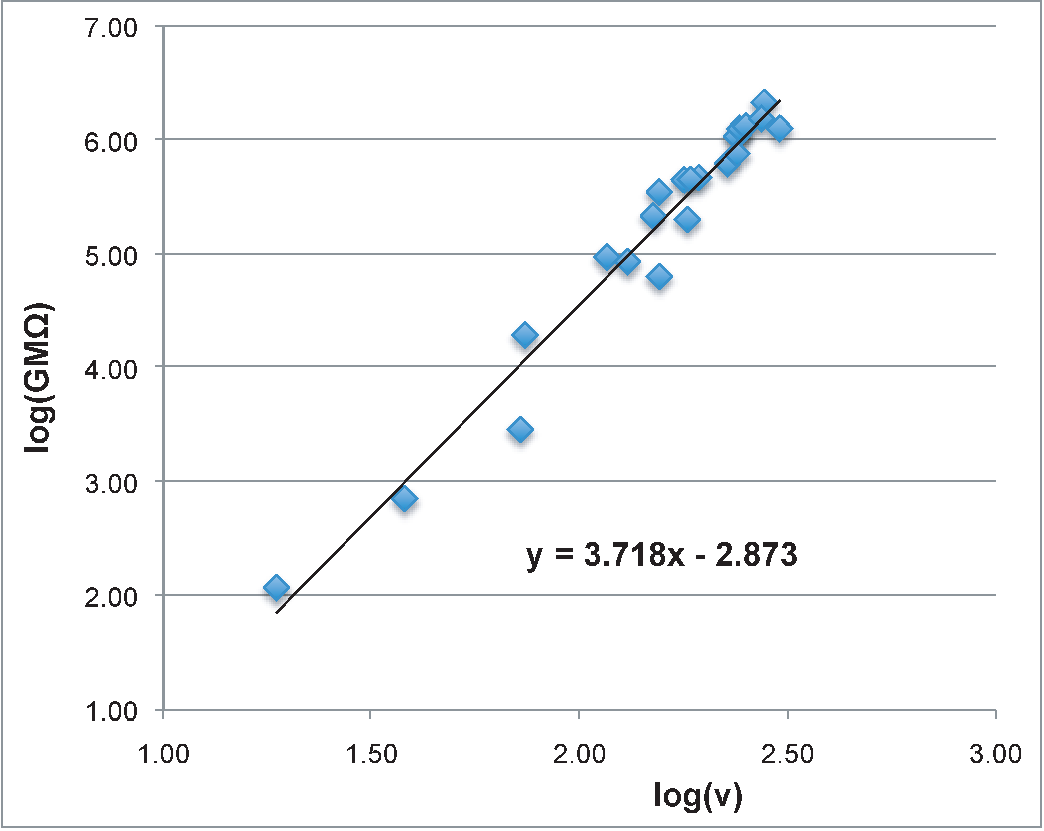}
\end{center}
\vspace{-0.5cm}
\caption{The luminous mass and velocity relation of the 22 galaxies fit  by the Tully-Fisher relation derived from the string model.}
\end{figure}

We  plot our best-fit values of $G\,M_{stellar}\, \Omega$ against $v$ in 
Fig.~\ref{fig:Tully-Fisher}.  
The representative velocity, $v$, is selected to be the {\it{maximal}} observed velocity in the 
entire curve for each galaxy, to eliminate man-made bias. 
 This no doubt introduces more scatter than necessary.
Despite that  the data obey the  relation remarkably well.  

Note that this is a nontrivial relation 
because  it relates  two parameters from two  additive force terms to an observed quantity, the
rotational speed.
Furthermore if one can determine the luminous mass of the galaxy, $M_{stellar}$ and the field strength,
$H$, we can determine the fundamental  charge-to-mass ratio. This ratio is universal for all  matter
according to string theory, and is determined by measuring the rotating speed of a test star.  Furthermore, our model
provides a dynamical explanation to the Tully--Fisher relation.

\subsubsection{A Relation Obtained from the String Model Fitting Result}
\label{subsec:relationtestof22galaxyfitting}
By dimensional analysis, together with some common results from astronomy, we can find a simple relationship between the field strength $\Omega$, galaxy luminosity and size. This serves as a consistency check for the string model. According to the string model, considering its analogy with electromagnetism, it is reasonable to expect the average field strength to be proportional to
$M^{\alpha} R^{-\beta}$,  where $M$ is the total luminous matter in the galaxy and $R$ is the size scale of the galaxy.~\footnote{By the Tully-Fisher relation~\citep{Tully&Fisher:1977} velocity is related to the total mass of the galaxy, therefore we do not need a separate term for the velocity dependence.} To be more specific we will heavily use the electromagnetism analogy in the following discussion. 
Consider a group of electrons azimuthal symmetrically distributed and
in rotation around the $z$ axis. 
Let us look at the magnetic field at the center of this distribution, the determining physical quantities are: the magnetic constant $\mu_0$, mass density scale $\rho_0$, distribution size scale $R_0$ and rotational angular velocity scale $\omega_0$.~\footnote{%
What we really want to check is the averaged field over the galaxy, but it is proportional to the field strength at the center.}%

\begin{figure}
\vspace{0.25cm}
\begin{center}
\includegraphics[width=80mm, height=70mm]{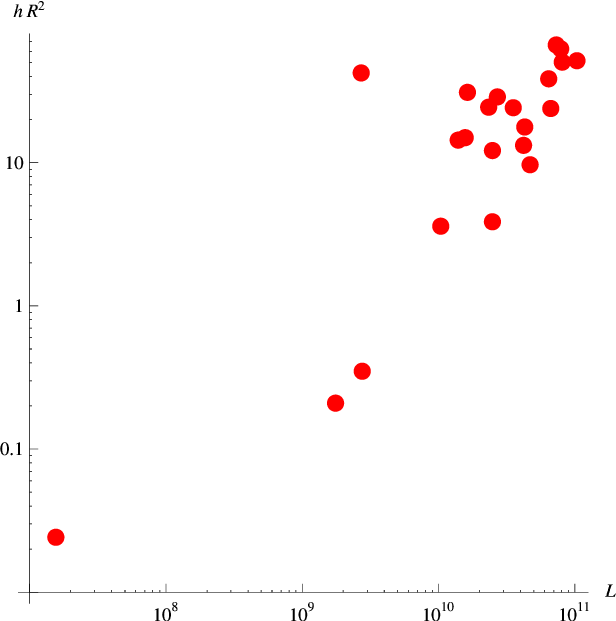}
\vspace{-0.25cm}       
\caption{luminosity to $\Omega R^2$ log-log plot}
\label{fig:L_hR2}       
\end{center}
\end{figure}

(Other determining factors include the shape and the spatial dependence of the mass distribution; and  the spatial distribution of the angular velocity. These factors do not  change the result of dimensional analysis but they do change the proportional constant.)
By dimensional analysis, we have 
\begin{equation}
  B\propto\mu_0\cdot \rho_0 \cdot\omega_0\cdot R_0^2
  \label{eq:dimensional_relation_B_and_other_factors_current_disk}
\end{equation}
Using the total charge  $Q\propto \rho_0\cdot R_0^3$, and defining $v_0=\omega_0\cdot R_0$, we have
\begin{equation}
  B\propto  \mu_0\cdot Q\cdot 
  R_0^{-2}\cdot v_0
  \label{eq:dimensional_relation_B_other_factors_variables_changed}
\end{equation}
Note that $v_0$ is the rotational velocity scale for the galaxy.  Translating to the language of the string model, it is
\begin{equation}
  \Omega\propto Q\cdot R_0^{-2}\cdot v_0
  \label{eq:dimensional_relation_Omega_1}
\end{equation}
The proportional constant here depends only on the mass and angular velocity distributions, or abstractly on the galaxy type.~\footnote{%
More precisely, it is not just the galaxy morphology type. The velocity distribution also matters.  It is possible that galaxies of the same morphology type but with very different velocity distributions will have different proportional constants.}
Thus galaxies with similar mass distribution profile and rotation curve \emph{shapes} should have similar constants of proportionality. 
Now recall $M\propto Q$, where the proportional constant is universal and thus the same for all galaxies. Furthermore we also use the assumption $L~\propto~M$.~\footnote{This relation is independent of the galaxy type. For more discussion about the mass luminosity-relation among galaxies of different types, see \citep{Roberts:1969}.}   Thus
\begin{equation}
  \Omega\propto L\cdot R_0^{-2} \cdot v_0
  \label{eq:dimensional_relation_OmegaR^2_Lv}
\end{equation}
To relate $v_0$ to $L$ we use the Tully-Fisher relation which says  $L~\propto~\Delta~V^{\alpha}$  where $\alpha$ is around $2.5\pm0.3$ and $\Delta V$ is the velocity width of the galaxy\citep{Tully&Fisher:1977}. 
The proportional constant in the Tully-Fisher relation is galaxy type independent.  Since $v_0$ is the overall scale for $\Delta V$, we  also have $L~\propto~(v_0)^\alpha$. 
Using this in (\ref{eq:dimensional_relation_OmegaR^2_Lv}), we get
\begin{equation}
  \Omega\cdot R_0^{2}\propto L^{1+\frac{1}{\alpha}}
  \label{eq:dimensional_relation_final_for_plotting}
\end{equation}
which after taking logarithm
\begin{equation}
  \ln\left( \Omega R_0^2 \right)=\left( 1+\frac{1}{\alpha} \right)
  \ln\left( L \right)+\ln\kappa
  \label{eq:dimensional_relation_log-log}
\end{equation}
where $\kappa$ is the proportional constant in the relation (\ref{eq:dimensional_relation_final_for_plotting}). 
The log-log diagram is shown in Fig.\ref{fig:L_hR2}.

There seems to be a trend of a linear relation in the ``main'' part of the diagram. 
The slope of the line is about $\frac{3}{2}$, quite close to the           $1+\frac{1}{2.5\pm0.3}$, which was derived theoretically. 
Two points seems to lie outside of the ``main'' part, i.e one at the lower left corner for the dwarf galaxy m81dwb, and another one for NGC4236 at the left of the upper right group. 
In terms of galaxy type these two galaxies are ``exotic'' among the 22, and thus perhaps their  $\ln\kappa$ deviated more from those in the ``main'' part. 
Actually in terms of morphology type, NGC4236 is SBdm, which is the most irregular among the regular types, while m81dwb is the only dwarf galaxy  among the 22 galaxies from the NED data base~\citep{NED, Mazzarella:2001wm}. All others are more or less regular galaxies. 
Presumably  these two lie on another line for irregular galaxies which is parallel to the line passing the  rest. 
It is possible that there are a series of parallel lines for different types of galaxies. 
However, statistical error from the small size of this data set makes the above arguments weak.  
Using analysis of this kind for a larger number and for more types of galaxies could make the situation clearer;  but this exercise is beyond the scope of the current project.  

\subsection{Discussion}
The original appeal of the NFW profile based on the ideas of hierarchical clustering was its universality.  
One simple NFW profile was expected to explain structure formation, rotation curves of galaxies--giant or dwarf--from high to low surface brightness.  
This promise has been undermined by the  cusp and core debate in dwarf galaxies  as well as  in the low surface brightness galaxies (see for example~\citep{vandenBosch:2000rza}).
The fact that light does not follow dark matter--well established by detailed observation and analysis in the Milky Way (see~\citep{Gilmore1997} for a summary),   in addition to  a clear deficit of satellite galaxies in MW  have only served to thicken the plot.  
This debate has recently  been taken to a broader context by observational  progress: The simplicity of the galaxies~\citep{Disney:2008jj} and the early assembly of the most massive galaxies~\citep{Collins:2009gz}  are at odd with the hierarchical clustering paradigm.  

While we are not claiming that our string toy model can answer all these questions in one stroke we merely show that it pans out  just as well as  the Dark Matter model in  fitting the galaxy  rotation curves while using  two fewer parameters.  
Moreover by tuning the ratio of the strength of the string field  to  stellar mass density   galaxies with a wide range of surface brightness and sizes can be accommodated.   
We have one dwarf and several LSB galaxies in our sample.
At the same time the model, based as it is on a tractable physical principle consistent with laws of mechanics and special relativity, does not suffer from the arbitrariness and puzzling inconsistencies of MOND. 

In order to describe  a universal galaxy rotation curve~\citep{Rubin:1985ze,Persic:1995ru,Salucci:2011ee} one  needs three parameters at most--to  specify the initial slope, where it bends, and the final slope.  
Any more parameter are redundant.  
In this regard the string model utilizes just the right number.  
The fact that it fits well on par with  the dark matter model which employs two extra parameters should be taken seriously.  
However one should guard against reading too much into the game of fitting.  
For example,  one cannot obtain a {{\it unique}} decomposition of the mass components of a galaxy using the rotation curve data alone, a difficulty encountered  in the context of  comparing different dark matter halo profiles.  
Acceptable fits (defined as $\chi < \chi_{min} +1$~\citep{Navarro:1998qc}) can be obtained with dark matter alone without any stellar matter in the CDM model.  
The roles of dark matter and stellar matter can also be completely reverted in the fitting routine.   
 On the other hand, the physical difference is dramatic.   
This  degeneracy is less severe in the string model in the sense that  the  string field cannot be completely traded off in favour of stellar matter, or  vice versa. 
However   a range of values for $R_{d}, \rho_{0}$ and $ H$,  where  ``acceptable'' fits can be obtained,  still exists.  
Therefore given the quality of the  available data  rotation curve fitting alone cannot distinguish between dark matter and  the string field in galaxies.

However precision measurements extended to   radii $r\sim 20R_{d}$ can  distinguish the string model from the other models: a  gently rising rotation curve  in this region is a signature prediction of this  string toy model.
At this moment we are, nevertheless, encouraged by this inchoate results to pursue further.  
In a separate article we shall subject our string model to other reality checks, and we shall report on how this simple string model accounts for gravitational lensing which is often cited  as  strong evidence for the existence of dark matter at intergalactic scales.

At this point it is worth mentioning that a critical reanalysis of available data performed by Kuijken and Gilmore  on velocity dispersion of F-dwarfs and K-giants in the solar  neighborhood,  with more plausible models concluded that  the data  provided no robust evidence for the existence of any missing mass associated with the galactic disk in the neighborhood of the
Sun~\citep{KuijkenGilmoreIII}.   
Instead a local volume density of  $\rho_0= 0.10M_{sun} pc^{-3}$ is favored,  which agrees with the value obtained  by star counting.  Dark matter would have to exist outside the galactic disk  in the form of  a gigantic halo. 
Their pioneer work was later corroborated  by~\citep{FlynnFuchs, Crezeetal, Pham, HolmbergFlynn, khriplovich2006upper} using  other sets of A-star, F-star and  G-giant data.  
Note that this observation can be nicely explained by our model as the field only affects the centripetal motion on the galactic plane; it has no   affect on the motion perpendicular to the galactic plane.  

We  presented a simple string toy model with only one free parameter and we  showed that  it can fit the galaxy rotation curves equally as well as the dark matter model with the  generalized NFW profile. 
The latter employs two more free parameters compared with the string model.   
The string model respects all known principles of physics and can be derived from the first principle using  string theory,  which in turn unifies gravity with other interactions. 
 Our model has an unambiguous prediction concerning the rotation dynamics of satellites and stars far away from the center of the (host) galaxy.   
 The ability to test the validity of string theory as a description of low energy physics makes the exercise worthwhile.

\section{Perihelion precession}
\label{sec:precession}
In this section we test the string model with planetary precession data  in the solar system. In~\citep{pitjeva2009epm,Iorio:2009}, 
after taking careful account of the influence of all other planets on the orbit of the planet concerned,
 anomalous precessions for Saturn were 
reported, for which no explanation within 
the standard paradigm seems to exist. 
This anomaly disappeared in successive analyses (Fienga et a. 2011, Pitjev \& Pitjeva 2013) in the sense that, nowadays, non-zero extra-precessions at a statistically significant level are absent; however intervals statistically compatible with zero for allowed values for any anomalous precessions are, indeed, obtained. 
It is thus an interesting laboratory to constrain our  string model and ask if the string gauge field can explain the reported ranges of possible anomalous precession.  
Furthermore,  this serves as an {\it independent} estimate of the upper bound on the strength of the string field in our galaxy. 
This estimate can thus be compared to the field strength estimate from     the Milky Way rotation curves. 
It is because we can use the reported ranges of anomalous precession to determine the  profile as well as a bound on the strength of the string gauge field inside the solar system.   
As it turns out, the extra field needed to generate the extra centripetal force to account for the anomalous precession has a profile of a dipole field generated by the Sun. 
It is then important to compare the magnitudes of  field strength as obtained  by different methods and observations.  A consistent model should give similar values in the field strength for the same object.

\subsection{Field Strength from Precession in the Solar System}
\label{subsec:precession_solar_system}

\begin{figure}
\begin{center}
\includegraphics[width=80mm, height=70mm]{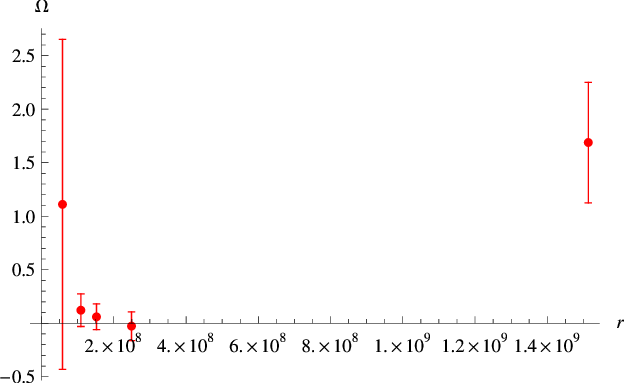}
\caption{$\Omega(1\times 10^{-17}s^{-1})$ versus distance
  $r(km)$ to the Sun}
\label{fig:omega_r}       
\end{center}
\end{figure}

Here we will use the anomalous precession data to determine the string field
profile and field strength in the solar system. We attribute all the anomalous precession  to the magnetic like force due to the string field.
We are interested in the quantity~\citep{CheungGRC0:2007}
\begin{equation} \label{eq:string_omega_definition}
\Omega=\frac{QH}{m}
 \end{equation}
where $\frac{Q}{m}$ and $H$ are the string charge-to-mass ratio and the field
strength, respectively.~\footnote{Note that here we dropped the $\frac{1}{2}$ factor in $\Omega$ as defined when discussing GRC fitting as constants of order one are not important.}  $\Omega$ has the  dimension, $s^{-1}$, that of frequency. 
In other words we are testing the validity of Newtonian gravity in the extremely low frequency regime.
The corresponding quantity  in electromagnetism is $\frac{eB}{m}$.
In these units it is easy for us to compare it with the strengths of other magnetic-like forces, e.g $\frac{eB}{m}$, 
in electromagnetism.
Precession from a magnetic-like force perturbation has been worked out in detail from the  first principle
in~\citep{xu2011perihelion} 
(see also~\citep{Ni:2012xu,adkins2007orbital,chashchina2008remark,ruggiero2010perturbations,D'Eliseo:2012js,Iorio:2010rk}.)
\begin{equation}
 \label{eq:precession_rate}
  \delta \dot{\omega}=-\left( \frac{qB}{m} \right)\frac{1}{\sqrt{GM_{\odot}}}\pi
  \left( \frac{a^{\frac{3}{2}}}{T} \right).
\end{equation}
To calculate the strength of $\Omega$ using precession data we  replace      $\frac{qB}{m}$ with    $\Omega=\frac{QH}{m}$ in the above formula~(\ref{eq:precession_rate}) and invert  it to get
\begin{equation}
  \Omega=(-\delta \dot{\omega})\sqrt{GM_{\odot}}\frac{T}{\pi
  a^{\frac{3}{2}}}.
  \label{eq:Omega_for_fitting}
\end{equation}
For error analysis, we have
\begin{eqnarray}\label{eq:error_Omega}
 \left|\frac{Err(\Omega)}{\Omega}\right|=\left|\frac{Err(\delta\dot{\omega})}{\delta\dot{\omega}}\right|+\frac{1}{2}\left|\frac{Err(GM_\odot)}{GM_\odot}\right|+\frac{3}{2}\left|\frac{Err(a)}{a}\right|.
\end{eqnarray}
However, relative errors from other sources are extremely small compared to $\frac{Err(\dot{\delta\omega)}}{\dot{\delta\omega}}$. Indeed, as can be seen from data in table \ref{tab:precession_rate_&_Omega}, $\frac{Err(\dot{\delta\omega)}}{\dot{\delta\omega}}$ is around order 1, $\frac{Err(a)}{a}\ll1$. Moreover, $GM_\odot=132712440042\pm10km^3/s^2$ (\citep{konopliv2011mars} page 425), i.e $\frac{Err(GM_\odot)}{GM_\odot}\sim 10^{-10}$. So we can compute the error by
\begin{equation}
  \left| \frac{Err(\Omega)}{\Omega}\right |=
  \left |\frac{Err(\delta\dot{\omega})}{\delta\dot{\omega}}\right|~.
  \label{eq:relative_error_equal_omega_and_Omega}
\end{equation}
The upper bounds of $\Omega$ determined from precession data are also shown in Table~\ref{tab:precession_rate_&_Omega},    and the fitting results are  shown in    Fig.~\ref{fig:omega_r}.    $a$, $T$ and $e$ are from HORIZON~\citep{HORIZON}. Uncertainties in $a$ are from \citep{pitjeva2007use} (Table 3). 

The central values of the allowed ranges of $\Omega$ for the inner planets exhibit a decreasing
pattern with respect to $r$, although big error bars also allow  for the case of the vanishing string field.~\footnote{The same argument applies to precession rate as well, which is proportional to $\Omega$.}
At Saturn, $\Omega$ is nonzero within one $\sigma$.
However, as mentioned in~\citep{pitjeva2009epm,Iorio:2009}, the error bar at Saturn may actually be bigger than quoted, in which case the value of precession, or $\Omega$, may vanish. More precise measurements on precession are needed to
definitely determine  the existence of $\Omega$  (in other words, the anomalous precession rate) in the solar system.
Inspired by this decreasing pattern in 
Appendix~\ref{subsec:profile_fitting} 
we will fit data of inner planets with a (nearly) power law profile.
No matter how critically we take the profile and magnitude  of $\Omega$ here,  it is--nevertheless--certain that the upper limit of $\Omega$ as deduced from the currently observed ranges of potential anomalous precession of planets in the solar system is on the order of $10^{-17}s^{-1}$,   but it can be zero as well.

\begin{table*}[!h]
\footnotesize
\begin{center}
\begin{tabular}{r||c|c||c|c||c|c}
Planet & $\delta\dot{\omega}$ (I09)& $\Omega$ & $\delta\dot{\omega}$ (INPOP08)& $\Omega$ & $\delta\dot{\omega}$ (INPOP10a) & $\Omega$ \\\tableline
Mercury & $-36\pm50$ & $1.11$ & $-100\pm300$ & $3.07$ & $4\pm6$ & $-0.123$ \\
Venus & $-4\pm5$ & $1.22\times10^{-1}$ & $-40\pm60$ & $1.22$ & $2\pm15$ & $-0.0608$ \\
Earth & $-2\pm4$ & $5.99\times10^{-2}$ & $0\pm2$ & 0 & $-2\pm9$ & $0.0599$\\
Mars & $1\pm5$ & $-2.69\times10^{-2}$ & $4\pm6$ & $-0.108$ & $-0.4\pm1.5$ & $0.0108$ \\
Saturn & $-60\pm20$ & 1.69 & $-100\pm80$ & 2.81 & $1.5\pm6.5$ & $-0.0422$  \\\tableline
Planet & $\delta\dot{\omega}$ (P09) & $\Omega$& $\delta\dot{\omega}$ (P13) & $\Omega$&&\\\tableline
Mercury& $-40\pm50$ & 1.23&$-20\pm30$&0.61&&\\
Venus& $240\pm330$ & $-7.30$&$26\pm16$&-0.79&&\\
Earth & $60\pm70$ & $-1.80$&$1.9\pm1.9$&-0.057&&\\
Mars& $-70\pm70$ & 1.88&$-0.2\pm0.37$&0.0054&&\\
Saturn& $-100\pm150$ & 2.81&$-3.2\pm4.7$&0.09&&\\
\end{tabular}
\caption{\label{tab:precession_rate_&_Omega}Anomalous precession rates ($10^{-4}~''/cy$) and corresponding upper bounds on $\Omega$ ($10^{-17}s^{-1}$). We do not show the uncertainties of $\Omega$ in the table, but they can be obtained from those of $\delta\dot{\omega}$ by the relation $\left| \frac{Err(\Omega)}{\Omega}\right |=
  \left |\frac{Err(\delta\dot{\omega})}{\delta\dot{\omega}}\right|$. Precession data I09 is from \citep{Iorio:2009}. INPOP08 and INPOP10a are from (\citep{fienga2011inpop10a},table 5). P09 is from (\citep{pitjeva2009epm}, table 8). P13 is from (\citep{pitjev2013constraints}, table 5).}
\end{center}
\end{table*}


As a comparison,  let us note that for the real magnetic field near the Earth,  $B\sim 10^{-9}~\mathrm{Tesla}$, and with $\frac{e}{m}=1.76\times 10^{11}C/kg$,  we have  $\Omega_{em}\sim 100s^{-1}$.
(See  Appendix~\ref{appendix_solarmagnetic} for details.)
In that sense, the string field strength is, naively,  $10^{19}$ times smaller than the strength of the  magnetic field near the earth.
We may wonder why this ``strong'' magnetic field has not affected precession of planets, and specifically, we can ask if it is related to the observed anomalous precession of planets.  One reason why we do not have to worry about  the real magnetic field is that planets are electrically neutral (See \citep{iorio2012constraining}), but  charged under the string gauge field as postulated. The real magnetic field can act on neutral matter only through dipole-dipole interaction, which, as explained in Appendix~\ref{appendix_solarmagnetic},  does not contribute to planet precession for several reasons.

There is however another question we may ask about the string model now that  we are assuming matter is  charged under the string field and being acted on by the corresponding magnetic part for this charge: is there an electrical part of the interaction between stringly charged matter? 
After all, we seem to be assuming that all matter take the ``same''  kind of charge. This question lies outside of our current model and calls for more  theoretical investigations into the nature of this string charge.  As for the model used here,  we are considering a magnetic interaction in the form of a Lorentz force~\footnote{%
It is amusing to discover  the extra acceleration due to the string field for objects on Earth. Since the field strength is acquired for the rest frame relative to the Sun, the velocity of objects on Earth should be nearly the same as the velocity of  Earth relative to the Sun, i.e $30km/s$. The corresponding acceleration produced is therefore $\sim 10^{-13}m\cdot s^{-2}$.}.


A few  words on  Saturn are warranted. One aspect special about Saturn is that it belongs to the gas giant group while all other planets considered here are small, solid, and belong to  the inner planet family of the solar system. 
As in the electromagnetic theory, the content and structures 
of planets may affect their interactions with the string field, which might explain the somewhat anomalous behavior of Saturn. This speculation could be supported if anomalous precession behaviors of other gas giants similar to  Saturn can be measured in the future.~\footnote{%
According to~\cite{pitjev2013constraints} the current perihelion precession measurement for Jupiter is consistent with zero as the statistical level of significance  is likely to be too small.%
} 
We await new data for Uranus, Neptune, and Pluto~\citep{2012IAUJD...7E..37P,2008IAUS..248...20P,1998srst.conf..207M} as well as data for Jupiter with increased accuracy from the JUNO~\citep{2007AcAau..61..932M,2004DPS....36.1401B}~\footnote{JUNO will also be used to test general relativity. See \cite{helled2011jupiter,anderson2004gravity,iorio2010juno,iorio2013possible}.} and JUICE~\citep{2012LPICo1683.1039D} missions.

We would also like to remark that there are  other approaches to addressing of the problems of anomalous precession of the  planets in the solar system--which are completely independent of ours and invoke different physics. (See~\citep{pitjeva2009epm,Fienga:2011qh,Turyshev:2009ir,Reynaud:2008yd} for an overview on how planetary dynamics can be used  as a probe for fundamental physics.)
The standard method  is to extrapolate galactic dark matter to the solar system to estimate its influence on planetary motion:~\citep{Khriplovich:2006uq,frere2008bound}.
For example, an array of models~\citep{leiva2012kepler,mirza2002noncommutative} use non-commutative geometry~(e.g.~\citep{cheung1998non,seiberg1999string}).
Another large class of models~\citep{Dvali:2002vf,Lue:2005ya,Iorio:2005ch,Iorio:2006ri,Gabadadze:2007as,Battat:2008bu,abdujabbarov2010test}
makes  use of   induced gravity~\citep{cheung2004open}
in  the brane-world a la DPG (see~\citep{lrr-2010-5,Gabadadze:2007dv} for nice reviews.).
Yet another way to explain the precession is based on modified newtonian dynamics, MOND,~\citep{Boyarsky:2010bd,Afshordi:2008rd,Schmidt:2008qi,Iorio:2009ke,Gabadadze:2007as,Iorio:2007rk,Iorio:2011gb,Iorio:2009fa}.
Effects from general relativity, as well as the modification of Einstein's gravity (see, for example~\citep{Sotiriou:2008rp,Clifton:2011jh,Hinterbichler:2011tt}, reviews on modified gravity),
are explored in  excruciating detail  to explain the anomalous precessions of planets.  The volume of literature is also large. A random sample of recent literature includes~\citep{Ni:2012xu, Damour:2008ji, Iorio:2012cm,Iorio:2012gr,Borka:2012tj,Lecian:2008vc,Reynaud:2008yd}).
And various other creative approaches: see, for example,~\citep{Iorio:2012pv,Arakida:2012ya,Gong:2009zzb,Kopeikin:2012iy,Iorio:2011aa,Iorio:2011vb,Iorio:2011zu,Iorio:2011jz,Iorio:2011zv,Iorio:2011ab}.

\subsection{Field strength from Milky Way rotation curve}
\label{subsec:GRC_milkyway}
In Section \ref{subsec:precession_solar_system} we obtained $\Omega$ in the solar system from anomalous precession. The constant background value, found by the profile fitting in Appendix~\ref{subsec:profile_fitting},  mainly comes from other matter in the Milky Way.
On the other hand, from  the same idea used in~\citep{CheungGRC1:2008}, we can also estimate $\Omega$ at the solar system due into the Milky way's rotation. Thus a natural check of the string model is to compare these two values of $\Omega$: one from the constant background from precession in the solar system and the other  one from the rotation curve of the Milky Way.

Directly using the idea in~\citep{CheungGRC1:2008}, we can make a rough estimate of $\Omega$ in the Milky Way as follows. 
In the string model, the total force on the galaxy mass is composed of only the gravitational attraction from visible mass in the Milky Way and the magnetic-like force from the string field. By increasing $r$, the gravitational force decreases quickly, and the magnetic like force always increases. At the position of the Sun the rotation curve is well into the  flat region~\citep{Sofue:1997,Sofue:1999jy,Sofue:2000jx,Battaner:2000ef}.
Therefore on the Sun the gravitational force should be negligible with respect to the magnetic-like force from the string field.
Then~\citep{CheungGRC0:2007}
\begin{equation}
  m \frac{v^2}{r}\approx QHv
  \label{eq:eom_centripedal_magnetic_force}
\end{equation}
and thus
\begin{equation}
  \Omega\approx \frac{v}{r}
  \label{eq:formula_estimate_Hfield_at_sun}
\end{equation}
For the sun~\citep{Sofue:1997,Sofue:1999jy,Sofue:2000jx,Battaner:2000ef}, $v_{\odot}\approx
200km\cdot s^{-1}$ and
$r_{\odot}\approx 7.6kpc$. So
\begin{equation}
  \Omega_{\odot}\approx 26s^{-1}\cdot\frac{km}{kpc}= 8.5\times10^{-16}s^{-1}
  \label{eq:result_Hfield_at_sun}
\end{equation}

This is the upper limit on $\Omega$ due to the Milky Way at the position of the Sun. Since there is still a portion of distance further out to nearly $20kpc$ where the curve is quite flat
(with velocity $\sim200km\cdot s^{-1}$)~\citep{Sofue:1997,Sofue:1999jy,Sofue:2000jx,Battaner:2000ef}, we could have used these distances instead of $7.6kpc$ and the value of $\Omega$ will be reduced by a factor of about three. In any case, it is safe to say the upper limit of $\Omega$ is on the order $10^{-16}s^{-1}$. This is the strength of the field component perpendicular to the galactic plane. To convert it to the solar
system, note that the north galactic pole and the north ecliptic pole form an angle of 60.2$^\circ$ (which means the field in the solar system would be reduced  by almost half), and the Milky Way rotates clockwise when viewed from the north galactic pole (which means the field is negative in the solar system since the planets  rotate \emph{counterclockwise} if viewed from the same direction). Therefore, from the rotation curve of the Milky Way the upper limit of the effective field strength in the solar system due to matter in the milky way is $-10^{-17}$ to $-10^{-16}s^{-1}$. This magnitude is close to the one found by direct precession calculation without the profile assumption, but it is not the case for the field direction.
The precession indicates that the field in the solar system is positive, while the  rotation curve of the Milky Way says it is negative. One possible explanation is that at places near the Sun, the field is dominated by the positive field  the Sun generates, and the Milky Way provides only the constant background in the solar system. The profile fitting of precession, apart from a dipole like part due to the Sun, indeed gives a negative background ($-0.02\times10^{-17}s^{-1}$). However the magnitude there is smaller by almost two orders of magnitude.

\subsection{Field Strength from the Double Pulsar}
Here we discuss if we can get a better constraint on $\Omega$ from the precession data of the double pulsar. The double pulsar PSR J0737--3039 is a binary system of two pulsars\citep{burgay2003increased,iorio2009prospects,kramer2006tests}. It is a highly relativistic system and thus has become a laboratory for tests of general relativity. Due to its relativistic nature, the orbits of the solar system's members  have  a huge perihelion (periastron) precession rate $\dot{\omega}=16.89947(68)$deg/yr (\citep{kramer2006tests}, Table 1). After subtracting the first  order post-Newtonian contribution from  precession, the remaining ``unexplained'' precession rate is (\citep{iorio2009prospects}, Equation~(14))
\begin{eqnarray}\label{eq:anomalous_precession_double_pulsar}
 \delta\dot{\omega}=\dot{\omega}_{exp}-\dot{\omega}_{1PN}=-0.00463\pm0.03233~\mathrm{deg/yr}.
\end{eqnarray}
This value of precession imposes a bound on the magnitude of $\Omega$ at the system's location. For the system, the orbit period $T=0.10225156248(5)$ day, the projected semi-major axis $a\sin i=1.415176(5)\mathrm{light~s}$, for which $i\approx90^\circ$ is the inclination angle, and the total mass is $2.85(2)M_\odot$ \citep{kramer2006tests,iorio2009prospects}. Here we only  consider  the order of magnitude of the constraint on $\Omega$, so we can use these values directly in the 
Equation~(\ref{eq:Omega_for_fitting}). The result is
\begin{eqnarray}\label{eq:Omega_from_double_pulsar}
 \Omega=-10431''\pm72840''/\mathrm{cy}=-(1.6\pm11.2)\times 10^{-11}s^{-1}.
\end{eqnarray}
This is a bound on $\Omega$ much higher than those from rotation curves and precessions of planets in the Solar system. Therefore the double pulsar does not give a better constraint on $\Omega$. Also note that the large uncertainty allows for the case $\Omega=0$.

\subsection{Discussion}

With the latest reported ranges of the possible anomalous precession of planets in the solar system, we obtained  upper limits on the field strength  of  the string  
gauge field--attributed to  universally  coupling  to all matter--at several places in the solar system. 
The maximal field strength allowed in the orbit of, say, Mercury is found to be on the order of $10^{-17}s^{-1}$.  
The distribution of the field in the solar system looks like a superposition of a constant background field 
 and a $r^{-3}$ decreasing ``dipole'' component in (\ref{eq:Omega_profile}).
We discussed a possible configuration where the Milky Way produced the constant  background  field and the Sun produced  the ``dipole'' component. A profile fitting is done for this configuration, with details provided in  Appendix A. The background is found to be negative,
which, when combined with analysis using electromagnetism analogy, correctly matches with the fact that the solar system and the Milky Way rotate in opposite directions. 

As a comparison with the field strength obtained  from precession in the solar system, we estimated the upper limit on the field strength of the string gauge field in the Milky Way directly from the rotation curve of the Milky Way at the position of the Sun.
This  estimate has an  order of magnitude similar to  that  from precession. However,  the direction is opposite.
If we consider the background-plus-dipole-field configuration used for profile fitting to be correct, then we should consider only the background value when comparing with the one estimated by the Milky Way rotation curve.
In that case, the background direction agrees with the one predicted by the Milky Way rotation curve, but the magnitude is smaller by 2 orders of magnitude.

\section{Summary}
We discussed the possibility that the rotational speed of the stars at  the center of a spiral galaxy is supported by the presence of string gauge field which couples universally to all forms of matter.  We compared the goodness of fit of the string model to that of the commonly accepted CDM  model with the generalized NFW profile for DM distribution~\footnote{The NFW profile with $\alpha=1$ is known to have difficulty fitting dwarf galaxies as well as galaxies with low surface brightness. Hence the generalized profile which leaves 
$\alpha$ a free parameter is called for.}.   
We fit rotational speed data of 22 spiral galaxies of varied size and luminosity.  DM  model fits marginally better ($9\%$)at the price of two more free parameters than the string model.
A  Tully--Fisher relation relating the visible mass and  field strength of the string gauge field  to velocity $ GM_{star}\Omega = v^{3}$ can be derived dynamically  from the string model, which is obeyed fairly well  by all 22 galaxies of varied sizes and luminosities.

The existence of the string gauge field is taken a step further and is applied to explain the currently reported ranges of potential anomalous precession of planets in the solar system.    
The extra field needed to generate the extra centripetal force  to account for the anomalous precession has  a profile of a dipole field generated by the Sun.

The values--or upper bound in the case of planetary precession--of string fields, which are 
\begin{itemize}
\item{galaxy rotation curves of a set of 22 galaxies (without the Milky way),}
\item{potential perihelion precession of some planets in the solar system and}
\item{the Milky Way rotation curve,}
 \end{itemize}
%
are summarized in Table~\ref{tab:observation_summary}.
\begin{table}
\begin{center}
  \begin{tabular}{c||cc|c}
\tableline
    & Solar System $\subset$ & Milky Way
    & 22 other galaxies\\\tableline
    Rotation Curve & & $|\Omega|\lesssim 10^{-16}$
    &$6\times10^{-18}\lesssim|\Omega|\lesssim10^{-15}$\\\tableline
    Precession &$|\Omega|\lesssim 0.61 \times 10^{-17}$ & &\\\tableline
  \end{tabular}
  \caption{%
Dimension for $\Omega$ is $s^{-1}$. Note: the first column indicates the kind of observation used, and the first row indicates the object observed.
}
\label{tab:observation_summary}
\end{center}
\end{table}
Interestingly--and also luckily for the string model--the results from these different methods lead to a similar order of magnitude for $\Omega$.

While supersymmetry is losing some of its lust because of is has not been detected  in LHC~\citep{Santanastasio:2013iz, Golling:2013bh, Delgado:2012eu, Chatrchyan:2013dsa, Chatrchyan:2012wa, ATLAS:2012ht, :2012tha,  Chatrchyan:2012rva, ATLAS:2011ad, Aad:2011hh},
 DM is also losing its most celebrated  candidate, 
 the lightest  supersymmetric neutral particle. (See, for example,~\citep{ArkaniHamed:2012gw, Ibanez:2013gf, Murayama:2012jh} for possible explanations of the no show.) 
 The game of explaining the  missing matter in the universe is becoming intriguing again.

\acknowledgments
We  thank Yuran Chen and Youhua Xu for their collaboration at the early stages of the precession project. 
We also thank Konstantin Savvidy for helpful discussions related to the rotation curve project. 
This work is partially funded  by the Priority Academic Program Development of Jiangsu Higher Education Institutions (PAPD).
The research done in this work has been  supported in parts by the National Science Foundation of China under the Grant No.~10775067  
as well as   
Research Links Programme of Swedish Research Council under contract No.~348-2008-6049.


\newcommand{\fig}[3]{\begin{figure}[htbp]\centering\includegraphics[width=#3\textwidth]{#1}\caption{#2}\label{fig-#1}\end{figure}}

\newcommand{\refig}[1]{Fig.\ \ref{fig-#1}}

\newcommand{\twofig[2]}{
\begin{align}
\includegraphics[width=0.48\textwidth]{#1} &
\includegraphics[width=0.48\textwidth]{#2}
\end{align}
}

\appendix
\section{Profile fitting of precession in solar system}
\label{subsec:profile_fitting}
The decreasing pattern of $\Omega$ for inner planets with respect to $r$ indicates it might be useful to fit these values with a power law term. Presumably we could attribute this $r$ dependent term to the Sun from which $r$ is measured. On the other hand, note that $\Omega$ at the Mars is negative, although it seems also sitting on the same curve passing the first three inner planets.

One possible configuration for $\Omega$ compatible with these 2 facts is then that, in addition to a power law term, there is also a weak constant background $\Omega$ with opposite direction to the $\Omega$ from the Sun.
The constant background might be provided by all other matter in the universe. The Milky Way should be the most important source of this influence.  Here we try to fit field strength at different inner planets with the following profile,
\begin{equation}
  \Omega(r)=A+B r^{-a}
  \label{eq:Omega_profile}
\end{equation}
For the actual fitting on computer, the profile used is
\begin{equation}
  \Omega(r)=\Omega_0+\Omega_1 \left( \frac{r}{10^7 km} \right)^{-a}
  \label{eq:Omega_profile_actual}
\end{equation}
The best fitting parameters for this profile is
\begin{align}
  \Omega_0 & = -0.0223787\times 10^{-17}s^{-1}\\\nonumber
  \Omega_1 & = 260.504\times 10^{-17}s^{-1}\\\nonumber
  a  &= 3.09953
  \label{eq:best_profile_fitting_parameters}
\end{align}
It is helpful to know the relative strength of the two components from the  Sun and the constant background, which is shown in 
Table~\ref{tab:relative_strength_table}.
\begin{table}
\begin{center}
  \begin{tabular}{l||c|c|c|c}
    \tableline
    Planet& Mercury & Venus & Earth & Mars \\
    \tableline
    ratio & 50.3 & 7.1 & 2.5 & 0.54 \\
    \tableline
  \end{tabular}
  \caption{\label{tab:relative_strength_table}relative strength: power law
term/constant term}
\end{center}
\end{table}
The power law term decreases with $r$ quickly relative to the background. We can safely say that in most areas in the solar system, the string field would just be around that background, which is on the order of $10^{-19}$Hz.
And that $a$ is found to be near 3, which is exactly the power for a dipole field. It means that the string field interaction between the sun and planets is similar to that between a magnetic dipole and charged particles.

Also note that the fitting result tells us that  the constant background in the solar system is negative. This is good news for the string model.
As in electromagnetism, we expect the string gauge field in a galaxy to be generated by the rotation of  matter (charged under the string gauge field) in the galaxy, just like rotating electric charge would generate a magnetic field. Since the Sun and planets in the solar system rotate in the opposite direction of that of stars' rotation in the Milky Way,  it is therefore reasonable to  expect the background field to be negative if we consider the one from the Sun as positive.
This is exactly what the profile fitting has  told us.
Here only data of the solar system was used, but  the conclusion is for the whole Milky Way, specifically for its rotation direction.

However, this conclusion should be taken with a grain of salt, as we will explain below.
Firstly, the long error bars of precession weaken the conclusion from this profile fitting. Secondly, there exists possibilities for the field from the Sun to be in the same direction with that of the Milky way. As in ordinary electromagnetic theory, for a right handed current disk the magnetic field is downward outside of the major current distribution, but upward if we go into the current disk somewhere, specifically at the center of the disk. There is a place within the concentration where the magnetic field changes its sign.~\footnote{Considering this, the $\Omega$
in~\citep{CheungGRC0:2007,CheungGRC1:2008} are position-averaged one over the galaxy.}
Thus even if the Milky Way is rotating in opposite direction from the solar system, if the Sun is too close inside the major mass concentration of the Milky Way, the background should still be positive. To our advantage, it is known that the Sun contains 99\% of the total mass in the solar system, so it is reasonable to assume all planets are well outside of most mass in the solar system. And   the solar system lies somewhat outside  the majority of mass concentration of the Milky Way.

\section{Solar system magnetic field and its effect on planet perihelion precession}
\label{appendix_solarmagnetic}

References for this appendix are\footnote{There are also studies on orbital motions of planets under the action of the Sun's electric charge \cite{iorio2012constraining,avalos2011precession}}~\cite{Parker, Parker:1958, Encyclopedia:1997, Meyer:Parker:Simpson:1956, Babcock:1961}. The Sun and most planets in the solar system have magnetic field due to dynamo effect. If we treat both the Sun and the planet as magnetic dipoles interacting in vacuum (which leads to a central force with $n=-4$), using data of magnetic fields of the Sun (around $1\sim 2$ gauss at the polar region) and the Earth  (around 0.6 gauss at the polar region), we can find the corresponding precession produced is nearly $10^{-6}$arcsec/cy, which is 2 orders of magnitude smaller than the observed one. In fact the magnetic field in the solar system is much more complicated than those produced by several dipoles in vacuum. First, the solar system is not empty but filled with particles emitted from the Sun, i.e the solar wind. Charged particles lock with it the magnetic field of the Sun and spread it all around in the solar system. From the Sun to about the position of the Earth, the magnetic force line is parallel to the radial stream of solar wind particle  and falls off by $r^{-2}$. From the position of the Earth to about position of the Mars is a field free region with $B<10^{-6}$ gauss. Further out to the position of the Jupiter is a region with disordered magnetic field with $B\sim 10^{-5}$ gauss. For precession, the most important feature of the solar system field is that it is oscillating. First, for the Sun the magnetic dipole axis is inclined relative to the rotational axis. This leads to an oscillating neutral current sheet. Therefore planets on the ecliptic plane is above the neutral current sheet for half of solar self rotation period, below for another half.
Since field directions above and below the neutral current sheet is opposite, the magnetic force experienced by the planets also change directions within one self rotation of the Sun. Second, the magnetic field of the Sun also changes direction every 22 years due to its differential rotation, which leads to another oscillation of magnetic field on the planets. Altogether these two oscillations make the magnetic field effect on precession neglectable with  respect to other accumulating effects.

\section{Different fitting models give different best fittings}
\label{appendix_fitting}

Fitting a galaxy consists of following steps:
\begin{itemize}
\item assume a density profile (a parametric description of the density),
\item adjust the values of free parameters to minimize an ``error function.''
\end{itemize}
The corresponding resulting parameters are called best fitting parameters. For a single galaxy we can define different error functions. It can be defined by rotation curve, by surface brightness or some other observation data. The point of this appendix is that we usually get different best fitting parameters when using different error functions. In particular, best fitting parameters for rotation curve are different from those for surface brightness. Below we provide a simple and idealized example to illustrate this point.
\paragraph{Real}
Suppose the real density distribution is a linearly decreasing function of radius and becomes zero outside of a cutoff radius,
\begin{eqnarray}
\rho(r)=\Bigl\lbrace
\begin{array}{cc}
\rho_0(1-\frac{r}{R}), & (0\leq r\leq R) \\
0. & (r\geq R)
\end{array}
\end{eqnarray}
where $\rho_0$ and $R$ are two fixed constants for this particular galaxy.~\footnote{Here we assume the galaxy is a disk and the distribution has only $r$ dependence. So it is actually more appropriate to call it surface density.}
The corresponding velocity, using Newtonian gravitation theory, is
\begin{eqnarray}
 v(r)=\Biggl\lbrace
\begin{array}{cc}
\sqrt{2\pi \rho_0 G_N (\frac{1}{2}-\frac{1}{3}\frac{r}{R})r},&(0\leq r\leq R),\\
\sqrt{2\pi \rho_0 G_N \frac{1}{6}\frac{R^2}{r}},&(r\geq R)
\end{array}
\end{eqnarray}
and brightness is (assuming light is proportional to mass) where $\gamma$ is light to mass ratio. Which gravity theory we use does not affect the conclusion of this appendix, so long as we use the same theory for any profile to derive the velocity.
\begin{eqnarray}
 B(r)=\Bigl\lbrace
\begin{array}{cc}
\gamma \rho_0 (1-\frac{r}{R}),&(0\leq r\leq R),\\
0,&(r\geq R)
\end{array}
\end{eqnarray}
\paragraph{Guessed}
Without knowing the real distribution, suppose we assumed for this galaxy a constant distribution profile
\begin{eqnarray}
 \rho(r)=\Bigl\lbrace
\begin{array}{cc}
\sigma, & (0\leq r\leq D) \\
0. & (r\geq D)
\end{array}
\end{eqnarray}
Here $\sigma$ and $D$ are two parameters, rather than constants, being fitted to get best fitting values, while $\rho_0$ and $R$ have particular fixed values for this galaxy. This density profile produces following velocity profile
\begin{eqnarray}
v(r)=\Biggl\lbrace
\begin{array}{cc}
\sqrt{\pi \sigma G_N r},&(0\leq r\leq D),\\
\sqrt{\pi \sigma G_N D^2\frac{1}{r}},&(r\geq D),
\end{array}
\end{eqnarray}
and brightness profile
\begin{eqnarray}
B(r)=\Bigl\lbrace
\begin{array}{cc}
\gamma \sigma,&(0\leq r\leq D)\\
0,&(r\geq D).
\end{array}
\end{eqnarray}

\paragraph{Fitting}
As said before, we can define different error functions to do the fitting.
We can use either rotation velocity or brightness for fitting. Ideally the error functions can be defined respectively for rotation velocity and surface brightness by
\begin{eqnarray}
&&\chi_v^2(D,\sigma)=\int_0^\infty dr'[v_{model}(r',D,\sigma)-v_{real}(r')]^2\\
&&\chi_B^2(D,\sigma)=\int_0^\infty dr'[B_{model}(r',D,\sigma)-B_{real}(r')]^2
\end{eqnarray}
After minimizing these two error functions we get two sets of best fitting parameters for $D$ and $\sigma$. Let's denote the best fitting value for velocity error function $\chi_v^2$ by $(D_A,\sigma_A)$, for brightness error function $\chi_B^2$ by $(D_B,\sigma_B)$. Below we show these two sets of values do not coincide.
\paragraph{}
If we use the error function for brightness, obviously the best fitting parameters are
\begin{equation}
 D_B=R,~\sigma_B=\frac{1}{2}\rho_0
\end{equation}
To show $(D_B,\sigma_B)\neq(D_A,\sigma_A)$ it is sufficient to show that $(D_B,\sigma_B)$ does not minimize $\chi_v^2$.
For simplicity we take units such that
\begin{eqnarray}
\rho_0=1,~~~R=1,~~~\gamma=1,~~~\pi G_N=1,
\end{eqnarray}
then
\begin{eqnarray}
D_B=1,~~~\sigma_B=\frac{1}{2}.
\end{eqnarray}
In this unit system the real velocity is
\begin{eqnarray}
 v(r)=\Biggl\lbrace
\begin{array}{cc}
\sqrt{\frac{r}{2}},&(0\leq r\leq 1),\\
\sqrt{\frac{1}{2r}},&(r\geq 1).
\end{array}
\end{eqnarray}
Taking 100000 as the upper limit of the integration, we find
\begin{eqnarray}
&&\chi_v^2(D_B=1,\sigma_B=0.5)=0.2,\\
&&\chi_v^2(D=0.9,\sigma=0.5)=0.05
\end{eqnarray}
Thus $(D_B,\sigma_B)\neq(D_A,\sigma_A)$. The best fitting parameters for brightness do not best fit the velocity curve.

\paragraph{Remarks}
Above we considered a simple and idealized galaxy and showed that best fitting parameters for different error functions are in general different, although we were doing the fitting for the same galaxy. Another point worth noting here is that if we had guessed at the correct density profile (which linearly decreases and vanishes beyond some cutoff radius), the two best fitting parameters will be the same. We get different fitting results because we used a ``wrong'' profile for this galaxy. In real life the mass distribution for a galaxy is extremely complicated and can not be exactly described by any simple ``profile function.''
Hence after we assume the profile, define the error function and then do the fitting, we will always get different fitting results for ``independent'' error functions (e.g velocity and brightness). If the guessed profile is closer to the real distribution we get closer results for fittings by different error functions. In other words, a big difference between fitting results by different error functions means the profile we guessed at is very different from the real one.

Given different profile assumptions for the galaxy distribution, we thus have a way to judge in some sense which one is more ``correct''. With each profile we can derive corresponding distributions of velocity, brightness and so on, and  with each of these distributions we can compute( if we have the data) an error function $\chi^2$, which is dependent on a set of parameters owned by this profile,
\begin{eqnarray}
 \mbox{Density Profile}\longrightarrow\Biggl\lbrace
\begin{array}{cc}\nonumber
\mbox{Velocity}&\chi^2_v\\
\mbox{Brightness}&\chi^2_B\\
\vdots&\vdots
\end{array}
\end{eqnarray}
If the profile perfectly match the real one, there exists a single set of parameters simultaneously making all $\chi^2$ zero. On the other hand, if the profile differs too much from the real one, even if we can make one of the $\chi^2$ small, the corresponding parameters will usually make other $\chi^2$'s very large. Therefore it makes sense to use, e.g. the average of several $\chi^2$'s as the error function to be minimized. This profile, with the corresponding minimizing parameters of the average error function, best describes the distribution averagely, i.e. considering distributions whose    $\chi^2$ is averaged.

However we cannot use this to \emph{find} the real distribution.
We can only compare profiles \emph{given} their profile assumptions and say which is better in describing velocity, brightness and so on, or if we use some averaged error function, also say which is better considering their general performances in describing various properties simultaneously.

\section{Detailed Fitting Results of 22 galaxies using dark matter model and using string model}
\label{appendix_RC}

In this appendix we present all the results on the data fitting.
Table~\ref{tab:galaxy_measurement} summarizes the relevant data for luminosity as well as for distance determination.  In Table~\ref{tab:best_fit_DM} and Table~\ref{tab:best_fit_string}  the best fit values for the five free parameters in the dark matter fitting and three free parameter in the string model fitting are presented.  Mass of the stellar mass and dark matter halos (in the case of dark matter model) as well as the mass to light ratios are computed from the best fit values for each galaxy.  These values are tabulated in
Table~\ref{tab:DM_mass_summary} (dark matter) and in
Table~\ref{tab:string_mass_summary} (string).
The rest of the appendix presents 22 graphs of dark matter fit and string fit side by side for each of the 22 galaxies.  In each of the graph small cubes with error bars represent observational data the other symbols, and the curves, represent theoretical values.  The X-axis is radius in $kpc$ and Y-axis is velocity  in $kms^{-1}$.

\begin{table}
\begin{center}
  \begin{tabular}{|l|c|c|c|}
  \tableline
galaxy&	B-magnitude&	mucin&	Distance(Mpc) \\\tableline \tableline
m81dwb&	-12.5&	N/A& 	3.5 \\ \tableline
ngc0628&	-20.60&	29.95&	9.77\\ \tableline
ngc0925&	-20.05&	29.85&	9.33\\ \tableline
ngc2403&	-19.56&	27.68&	3.44\\ \tableline
ngc2976&	-18.12&	28.10&	4.17\\ \tableline
ngc3031&	-21.54&	28.57&	5.18\\ \tableline
ngc3184&	-19.88&	30.20&	10.96\\ \tableline
ngc3198&	-20.44&	30.42&	12.13\\ \tableline
ngc3521&	-21.08&	30.29&	11.43\\ \tableline
ngc3621&	-20.51&	29.58&	8.24\\ \tableline
ngc3938&	-20.01&	30.81&	14.52\\ \tableline
ngc4236&	-18.10&	27.08&	2.61\\ \tableline
ngc4321&	-22.06&	31.90&	23.99\\ \tableline
ngc4536&	-21.79&	32.11&	26.42\\ \tableline
ngc4569&	-21.10&	30.52&	12.71\\ \tableline
ngc4579&	-21.68&	31.80&	22.91\\ \tableline
ngc4625&	-17.63&	30.35&	11.75\\ \tableline
ngc4725&	-21.76&	31.45&	19.50\\ \tableline
ngc5055&	-21.20&	30.09&	10.42\\ \tableline
ngc5194&	-20.51&	30.01&	10.05\\ \tableline
ngc6946&	-20.89&	29.12&	6.67\\ \tableline
ngc7331&	-21.58&	30.75&	14.13\\ \tableline
\end{tabular}
    \caption{%
Relevant galaxy observation data for determining luminosity and distance.}
\label{tab:galaxy_measurement}
\end{center}
\end{table}


\begin{table}
\begin{center}
  \begin{tabular}{|l|c|c|c|c|c|c|}
  \tableline
galaxy& likelihood&	$R_{d}(kpc)$&	$\frac{r_{s}}{R_{d}}$& $\alpha$ &$\rho$& $\sigma$ \\\tableline \tableline
m81dwb&	0.158	&0.24&	8.2&	1.54	&613	&19.2\\ \tableline
ngc0628	&4.398	&1.88&	9.9	&0.75	&8409&	98.4 \\ \tableline
ngc0925	&2.392	&0.79&	27.6&	0.22&	859&	56.6 \\ \tableline
ngc2403	&4.467	&0.75&	21.1&	0.76&	744&	84.6\\ \tableline 
ngc2976	&0.504	&2.52&	18.3&	0.38&	998&	16.8\\ \tableline
ngc3031	&5.614	&3.99&	5.9	& 0.96	&705	&57.7\\ \tableline
ngc3184	&0.573	&3.97&	6.1	&0.58	&784	&42.1\\ \tableline
ngc3198	&0.441	&0.51&	18.2&	1.09&	291	&99.95\\ \tableline
ngc3521	&0.167	&0.57&	18.8&	1.56&	1738&	99.5\\ \tableline
ngc3621	&0.125	&6.85&	1.8	&1.80	&903	&5.6\\ \tableline
ngc3938	&0.905	&2.45&	5.5	&1.33	&747	&60.6\\ \tableline
ngc4236	&1.180	&4.40&	6.5	&0.83	&603	&1.4\\ \tableline
ngc4321	&0.956	&0.87&	19.6&	1.02&	1581&	80.5\\ \tableline
ngc4536	&0.598	&0.99&	10.7&	0.86&	105	&97.2\\ \tableline
ngc4569	&0.651	&0.74&	18.1&	0.95&	1350&	145.4\\ \tableline
ngc4579	&0.719	&7.90&	4.3	&   1.80&   1200& 3.4\\ \tableline
ngc4625	&0.874	&1.91&	14.1&	0.20&   191&	42.6\\ \tableline
ngc4725	&0.481	&4.00&	7.5	&0.77	&225	&53.9\\ \tableline
ngc5055	&1.133	&2.25&	4.5	&1.66	&2044	&67.0\\ \tableline
ngc5194	&0.442	&1.45&	4.4	&1.50	&1518	&49.9\\ \tableline
ngc6946	&5.670	&2.62&	29.4&	0.27&	2491&	42.6\\ \tableline
ngc7331	&0.839	&0.50&	19.6&	1.32&	1325&	197.7\\ \tableline
  \end{tabular}
  \caption{%
Best fit values for the five free parameters in the Dark Matter model with generalized NFW profile.}
\label{tab:best_fit_DM}
\end{center}
\end{table}

\begin{table}
\begin{center}
  \begin{tabular}{|l|c|c|c|c|}
  \tableline
galaxy&	likelihood&	$\Omega$(Hz$\cdot$km$/$kpc)& $\rho$&	$R_d$(kpc) \\\tableline \tableline
m81dwb&	0.125&	1.119&	15919&	0.15\\ \tableline
ngc0628&	4.347&	8.872&	11738&	1.80\\ \tableline
ngc0925&	2.452&	5.077&	500&	2.47\\ \tableline
ngc2403&	4.247&  13.418&	16497&	0.52\\ \tableline
ngc2976&	0.400&	0.100&	2135&	1.87\\ \tableline
ngc3031&	5.698&	2.000&	3187&	4.39\\ \tableline
ngc3184&	0.574&	0.655&	1552&	4.69\\ \tableline
ngc3198&	0.275&	1.968&	1971&	3.52\\ \tableline
ngc3521&	0.588&	6.046&	26678&	1.48\\ \tableline
ngc3621&	0.366&	10.338&	7413&	1.08\\ \tableline
ngc3938&	1.030&	9.722&	16199&	1.24\\ \tableline
ngc4236&	0.325&	10.358&	110&	2.02\\ \tableline
ngc4321&	2.070&	5.208&	3752&	3.15\\ \tableline
ngc4536&	0.739&	1.465&	734& 	5.87\\ \tableline
ngc4569&	0.662&	16.370&	12262&	1.04\\ \tableline
ngc4579&	0.688&	7.385&	5164&	3.00\\ \tableline
ngc4625&	0.548&	0.100&	1140&	1.45\\ \tableline
ngc4725&	0.268&	1.378&	1523&	6.73\\ \tableline
ngc5055&	3.521&	4.064&	24581&	1.54\\ \tableline
ngc5194&	0.807&	3.186&	10718&	1.10\\ \tableline
ngc6946&	6.035&	7.440&	4980&	1.80\\ \tableline
ngc7331&	0.522&	8.465&	18613&	1.68\\ \tableline
\end{tabular}
    \caption{%
Best fit values for the three free parameters in the string model.}
\label{tab:best_fit_string}
\end{center}
\end{table}

%
%
%
%
%
\begin{table}
\begin{center}
  \begin{tabular}{|l|c|c|c|c|c|c|}
  \tableline
galaxy&	B-magnitude&	 M$^{*}$&	DM+M*&	 M/L&	log(M/L)&	M*/DM\\\tableline \tableline
m81dwb&	-12.5&	4.07E+06&	7.89E+08&	50.72&	1.71&		3.12\% \\\tableline
ngc0628&	-20.60&	2.74E+10&	3.34E+12&	123.36&	2.09&		1.00\% \\\tableline
ngc0925&	-20.05&	2.08E+08&	2.26E+12&	138.71&	2.14&		0.16\% \\\tableline
ngc2403&	-19.56&	1.55E+08&	1.70E+12&	164.12&	2.22&		0.07\% \\\tableline
ngc2976&	-18.12&	7.79E+09&	4.79E+12&	1738.70&	3.24&		0.14\% \\\tableline
ngc3031&	-21.54&	2.17E+10&	3.83E+12&	59.63&	1.78&		3.42\% \\\tableline
ngc3184&	-19.88&	2.39E+10&	2.46E+12&	176.74&	2.25&		3.16\% \\\tableline
ngc3198&	-20.44&	1.88E+07&	4.71E+11&	20.18&	1.30&		8.90\% \\\tableline
ngc3521&	-21.08&	1.54E+08&	8.70E+11&	20.68&	1.32&		4.83\% \\\tableline
ngc3621&	-20.51&	1.41E+11&	1.94E+11&	7.81&	0.89&		2.37\% \\\tableline
ngc3938&	-20.01&	5.35E+09&	8.86E+11&	56.45&	1.75&		1.70\% \\\tableline
ngc4236&	-18.10&	2.51E+10&	7.30E+10&	27.01&	1.43&		0.61\% \\\tableline
ngc4321&	-22.06&	5.11E+08&	2.20E+12&	21.25&	1.33&		2.58\% \\\tableline
ngc4536&	-21.79&	4.96E+07&	6.24E+11&	7.71&	0.89&		11.57\% \\\tableline
ngc4569&	-21.10&	2.67E+08&	2.09E+12&	48.82&	1.69&		0.32\% \\\tableline
ngc4579&	-21.68&	2.88E+11&	8.96E+11&	12.26&	1.09&		7.57\% \\\tableline
ngc4625&	-17.63&	6.45E+08&	2.95E+12&	1683.06&	3.23&		0.06\% \\\tableline
ngc4725&	-21.76&	7.00E+09&	6.76E+12&	85.89&	1.93&		3.34\% \\\tableline
ngc5055&	-21.20&	1.13E+10&	5.00E+11&	10.64&	1.03&		8.81\%\\\tableline
ngc5194&	-20.51&	2.25E+09&	8.27E+10&	3.32&	0.52&		8.44\%\\\tableline
ngc6946&	-20.89&	2.19E+10&	7.09E+13&	2006.87&	3.30&		0.02\%\\\tableline
ngc7331&	-21.58&	8.09E+07&	1.33E+12&	19.92&	1.30&		3.24\%\\\tableline
\end{tabular}
    \caption{%
Summary of stellar mass and mass of dark matter halos computed from the best fit values, as well as mass to light ratios of the 22 galaxies in the dark matter model.  The mass is in units of $M_{sun}$. Mass to light ratio is relative to the mass of light ratio of the Sun.}
\label{tab:DM_mass_summary}
\end{center}
\end{table}

\begin{table}
\begin{center}
  \begin{tabular}{|l|c|c|c|c|}
  \tableline
galaxy&	B-magnitude&	$\Omega$(Hz$\cdot$km/kpc)&	total mass($M_{sun}$)&	 mass/light
 \\\tableline \tableline
m81dwb&	-12.5&		1.1193&	2.46E+07&	1.5814\\\tableline
ngc0628&	-20.60&	8.8716&	3.35E+10&	1.2388\\\tableline
ngc0925	&-20.05&	5.08&	3.68E+09&	0.2261\\\tableline
ngc2403	&-19.56&	13.4183&	1.12E+09&	0.1076\\\tableline
ngc2976	&-18.12&	0.10&	6.76E+09&	2.4549\\\tableline
ngc3031	&-21.54&	2.00&	1.31E+11&	2.0420\\\tableline
ngc3184	&-19.88&	0.6553&	7.77E+10&	5.5805\\\tableline
ngc3198	&-20.44&	1.9685&	4.19E+10&	1.7963\\\tableline
ngc3521	&-21.08&	6.0462&	4.20E+10&	0.9990\\\tableline
ngc3621	&-20.51&	10.3377&	4.60E+09&	0.1848\\\tableline
ngc3938	&-20.01&	9.7216&	1.50E+10&	0.9577\\\tableline
ngc4236	&-18.10&	10.3577&	4.42E+08&	0.1636\\\tableline
ngc4321	&-22.06&	5.2077&	5.70E+10&	0.5492\\\tableline
ngc4536	&-21.79&	1.4650&	7.22E+10&	0.8919\\\tableline
ngc4569	&-21.10&	16.37&	6.75E+09&	0.1575\\\tableline
ngc4579	&-21.68&	7.39&	6.79E+10&	0.9286\\\tableline
ngc4625	&-17.63&	0.10&	1.68E+09&	0.9565\\\tableline
ngc4725	&-21.76&	1.3785&	2.26E+11&	2.8680\\\tableline
ngc5055	&-21.20&	4.0642&	4.40E+10&	0.9373\\\tableline
ngc5194	&-20.51&	3.1862&	6.98E+09&	0.2803\\\tableline
ngc6946	&-20.89&	7.4403&	1.43E+10&	0.4037\\\tableline
ngc7331	&-21.58&	8.4646&	4.30E+10&	0.6446\\ \tableline
\end{tabular}
    \caption{%
Summary of stellar mass as well as mass to light ratios as computed from the best fit values in string model. The mass of the galaxies is in units of $M_{sun}$. Mass to light ratio is relative to the mass of light ratio of the Sun.}
\label{tab:string_mass_summary}
\end{center}
\end{table}

\begin{figure}[htbp]
\begin{align*}
\includegraphics[width=0.45\textwidth]{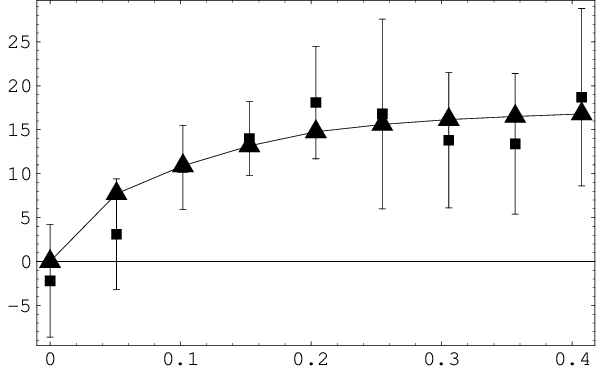}
 &~~
\includegraphics[width=0.45\textwidth]{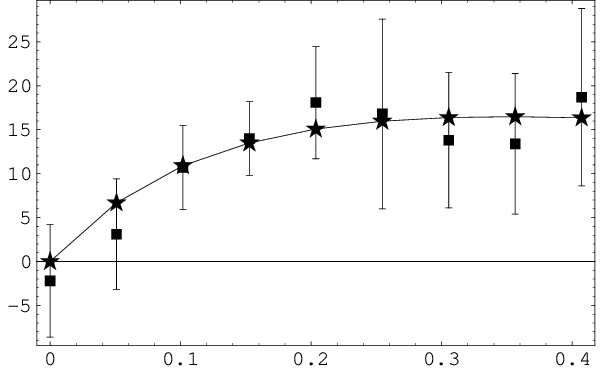}
\end{align*}
\caption{Rotation curve of dwarf galaxy m81dwb fit with the dark matter model (left) and with the  string model (right).
The $\chi$-squared value per degree of freedom using the dark matter model with a NFW profile is
$0.158$ while that using  the string model is $0.1254$.
}
\end{figure}


\begin{figure}[htbp]
\begin{align*}
\includegraphics[width=0.40\textwidth]{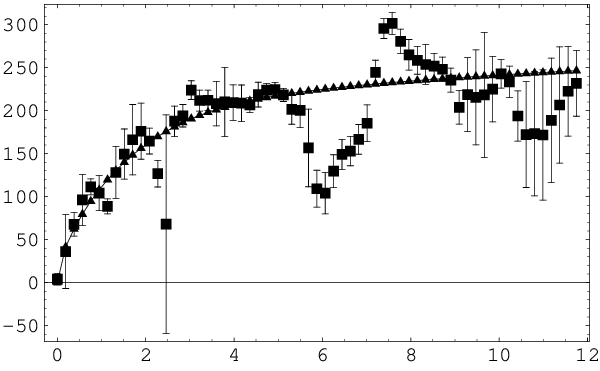}
 &~~
\includegraphics[width=0.40\textwidth]{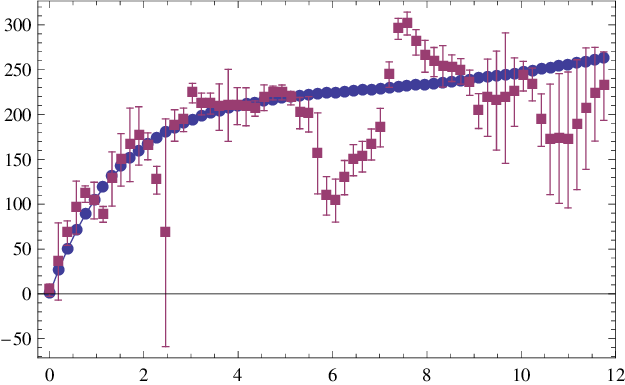}
\end{align*}
\caption{%
Rotation curve of NGC~0628 fit with the dark matter model   (left) and with the  string model  (right).
The $\chi$-squared value per degree of freedom using the dark matter model with a NFW profile is $4.398$ while that using  the string model is $4.3468$.%
}
\end{figure}

\clearpage

\begin{figure}[htbp]
\begin{align*}
 \includegraphics[width=0.40\textwidth]{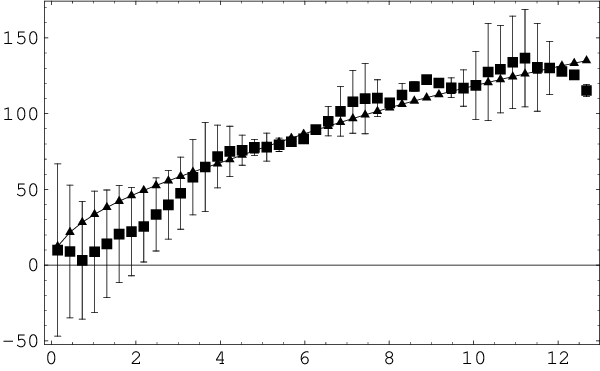}
&~~
  \includegraphics[width=0.40\textwidth]{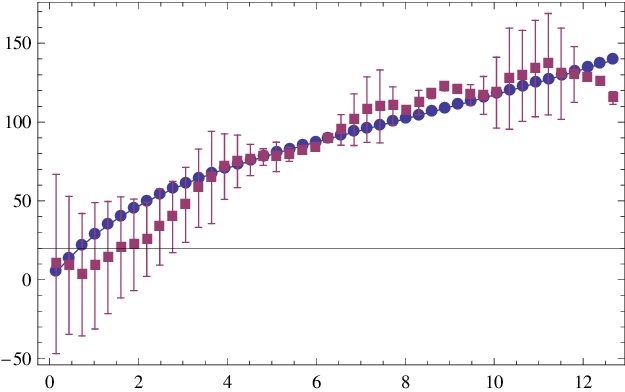} 
\end{align*}
\caption{Rotation curve of ngc~0925 fit with the dark matter model   (left) and with the  string  model  (right).
The $\chi$-squared value per degree of freedom using the dark matter model with a NFW profile is
$2.392$ while that using  the string model is $2.4518$.   }
\end{figure}


\begin{figure}[htbp]
\begin{align*}
\includegraphics[width=0.45\textwidth]{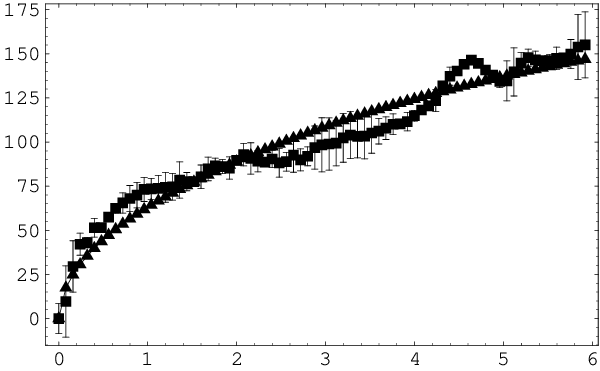}
 &~~~
\includegraphics[width=0.45\textwidth]{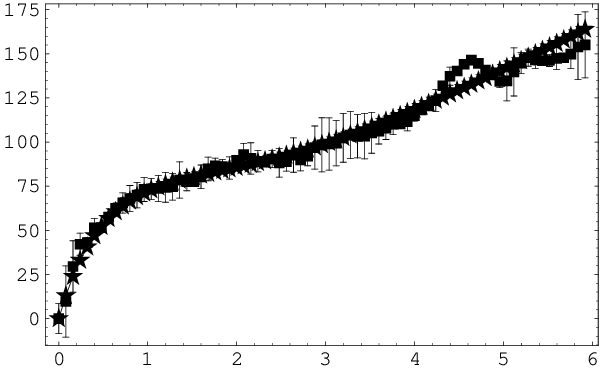}
\end{align*}
\caption{Rotation curve of NGC2403 fit with the dark matter model   (left) and with the  string  model  (right).
The $\chi$-squared value per degree of freedom using the dark matter model with a NFW profile is $4.467$ while that using  the string model is $4.2468$.
}
\end{figure}


\begin{figure}[htbp]
 \begin{align*}
  \includegraphics[width=0.48\textwidth]{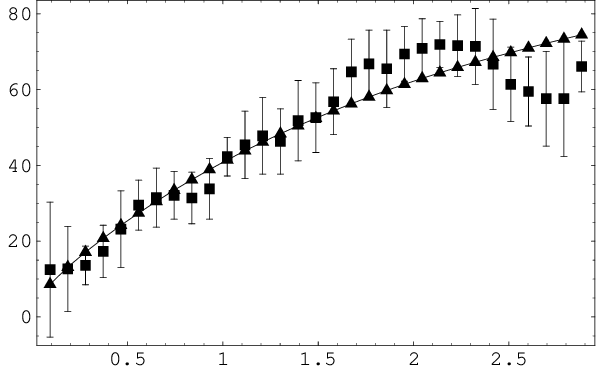}
 &
  \includegraphics[width=0.48\textwidth]{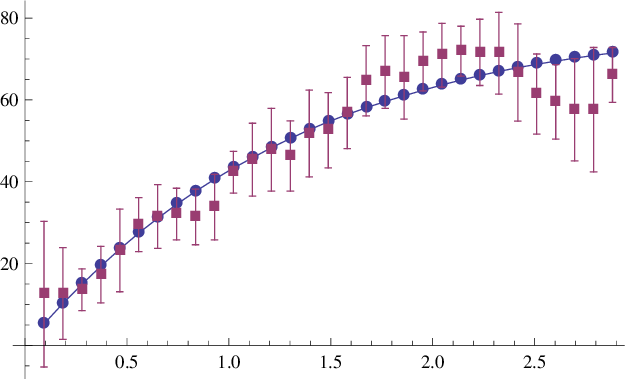}
\end{align*}
\caption{ Rotation curve of ngc~2976 fit with the dark matter model   (left) and with the  string model  (right).
The $\chi$-squared value per degree of freedom using the dark matter model with a NFW profile is $0.504$ while that using  the string model is $0.4001$.
}
\end{figure}

\begin{figure}[htbp] \label{fig:ngc3031}
\begin{align*}
\includegraphics[width=0.45\textwidth]{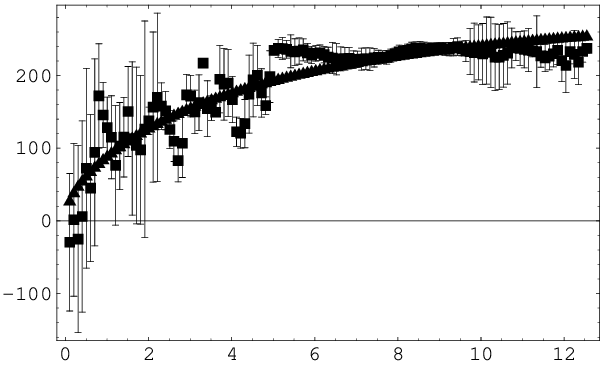}
 &~~
\includegraphics[width=0.45\textwidth]{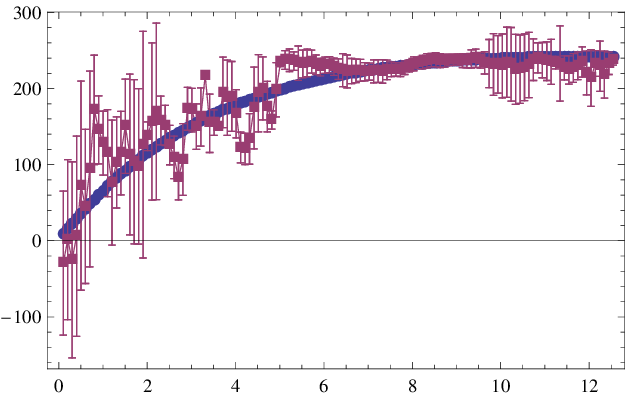}
 \end{align*}
\caption{
Rotation curve of ngc~3031 fit with the dark matter model   (left) and with the  string model  (right).   The $\chi$-squared value per degree of freedom using the dark matter model with a NFW profile is $5.614$ while that using  the string model is $5.6977$.   }
\end{figure}


\begin{figure}[htbp]
\begin{align*}
\includegraphics[width=0.45\textwidth]{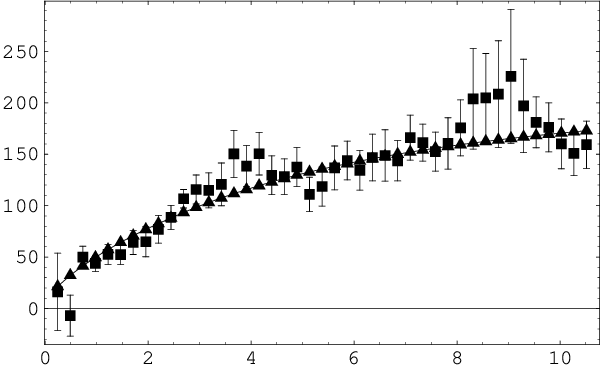}
 &~~
\includegraphics[width=0.45\textwidth]{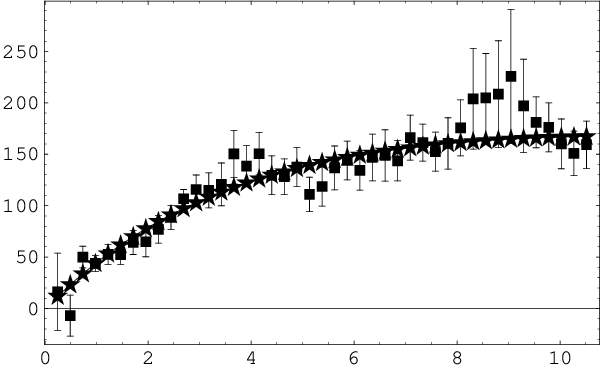}
\end{align*}
\caption{Rotation curve of NGC~3184 fit with the dark matter model   (left) and with the  string model  (right).
The $\chi$-squared value per degree of freedom using the dark matter model with a NFW profile is
$0.573$ while that using  the string model is $0.5740$.
}
\label{fig:string_ngc3184}
\end{figure}

\begin{figure}[htbp]
\begin{align*}
\includegraphics[width=0.45\textwidth]{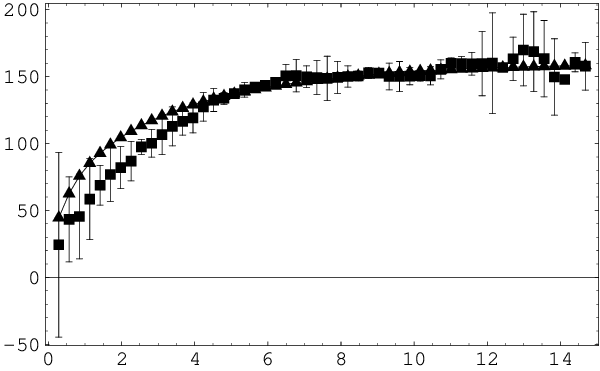}
 &~~
\includegraphics[width=0.45\textwidth]{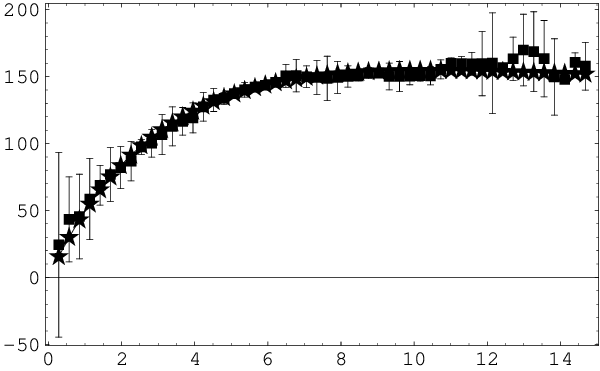}
\end{align*}
\caption{Rotation curve of ngc~3198 fit with the dark matter model   (left) and with the  string model  (right).
The $\chi$-squared value per degree of freedom using the dark matter model with a NFW profile is
$0.441$ while that using  the string model is $0.2748$.   }
\label{fig:ngc3198}
\end{figure}

\begin{figure}[htbp]
\begin{align*}
\includegraphics[width=0.45\textwidth]{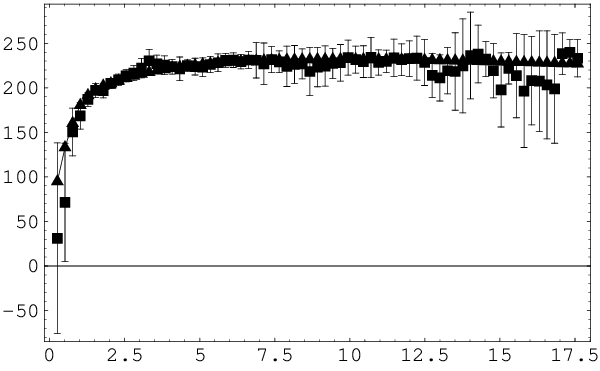}
 &~~
\includegraphics[width=0.45\textwidth]{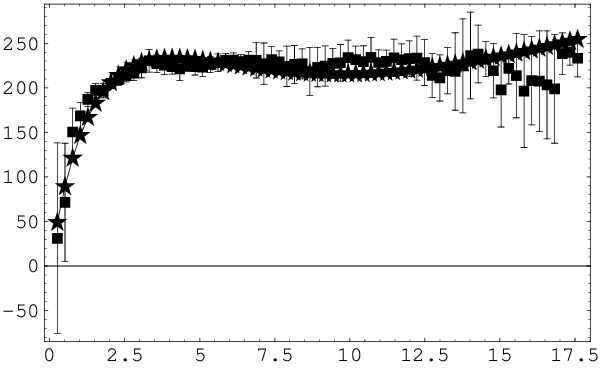}
\end{align*}
\caption{Rotation curve of ngc~3521 fit with the dark matter model   (left) and with the  string
model  (right).
The $\chi$-squared value per degree of freedom using the dark matter model with a NFW profile is
$0.167$ while that using  the string model is $0.5875$.}
\end{figure}

\begin{figure}[htbp]
\begin{align*}
\includegraphics[width=0.45\textwidth]{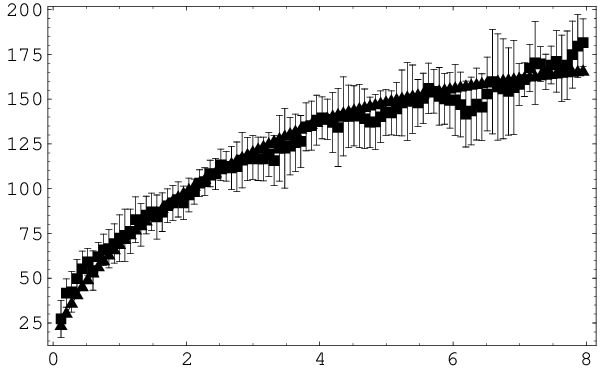}
 &~~
\includegraphics[width=0.45\textwidth]{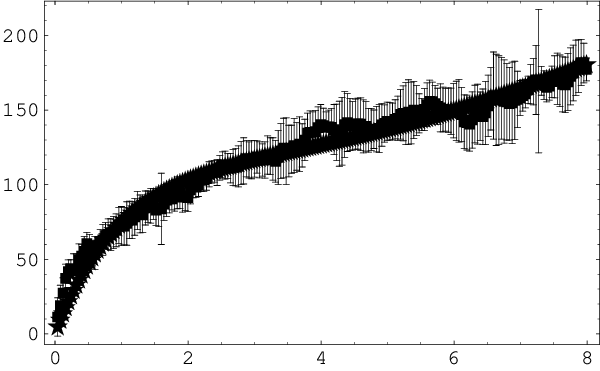}
\end{align*}
\caption{Rotation curve of NGC3621 fit with the dark matter model   (left) and with the  string  model  (right).
The $\chi$-squared value per degree of freedom using the dark matter model with a NFW profile is
$0.125$ while that using  the string model is $0.3658$.}
\label{fig:string_ngc3621}
\end{figure}

\begin{figure}[htbp]
\begin{align*}
\includegraphics[width=0.45\textwidth]{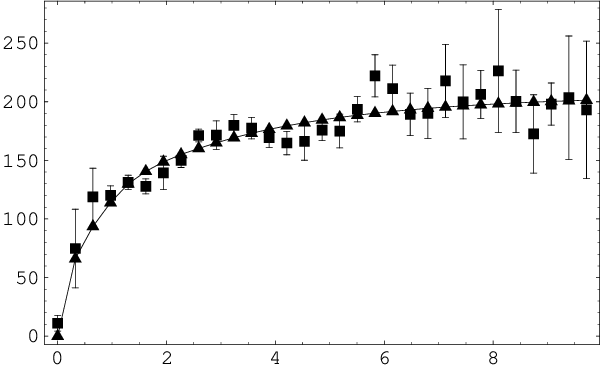}
 &~~
\includegraphics[width=0.45\textwidth]{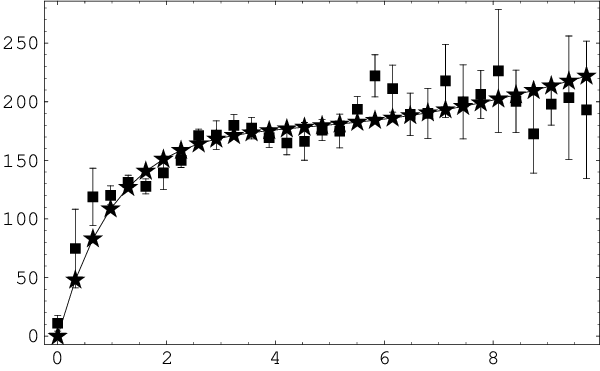}
\end{align*}
\caption{Rotation curve of ngc~3938 fit with the dark matter model   (left) and with the  string   model  (right).
The $\chi$-squared value per degree of freedom using the dark matter model with a NFW profile is
$0.905$ while that using  the string model is $1.0304$.
}
\label{fig:ngc3938}
\end{figure}

\begin{figure}[htbp]
\label{fig:ngc4236}
\begin{align*}
\includegraphics[width=0.45\textwidth]{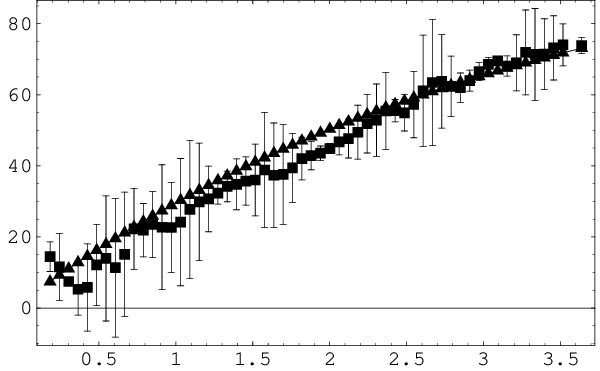}
 &~~~
\includegraphics[width=0.45\textwidth]{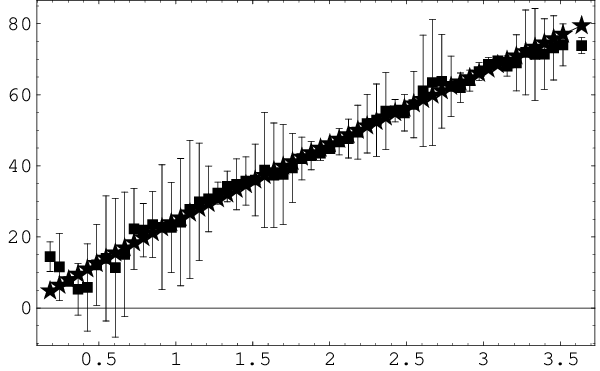}
\end{align*}
\caption{Rotation curve of ngc~4236 fit with the dark matter model   (left) and with the  string  model  (right).
The $\chi$-squared value per degree of freedom using the dark matter model with a NFW profile is
$1.180$ while that using  the string model is $0.3255$.   }
\end{figure}

\begin{figure}[htbp] \label{fig:ngc4321}
\begin{align*}
  \includegraphics[width=0.45\textwidth]{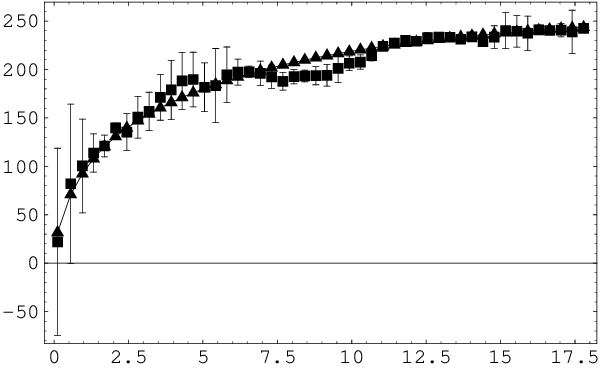}
 &~~~
  \includegraphics[width=0.45\textwidth]{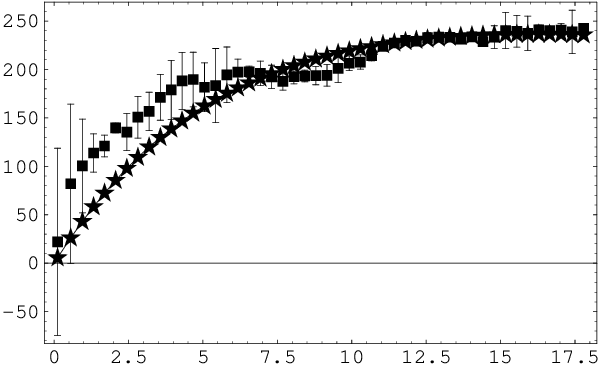}
\end{align*}
\caption{Rotation curve of NGC4321 fit with the dark matter model   (left) and with the  string
model  (right).
The $\chi$-squared value per degree of freedom using the dark matter model with a NFW profile is
$0.956$ while that using  the string model is $2.0703$.   }
\end{figure}

\begin{figure}[htbp]
\begin{align*}
\includegraphics[width=0.45\textwidth]{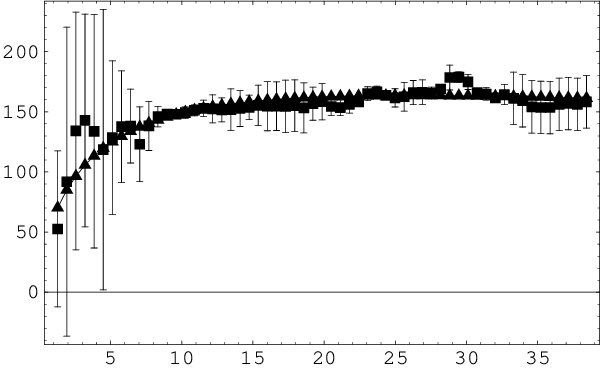}
 &~~~
\includegraphics[width=0.45\textwidth]{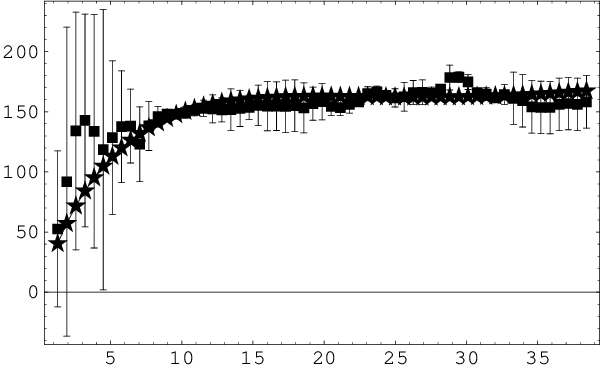}
\end{align*}
\caption{Rotation curve of NGC4536 fit with the dark matter model   (left) and with the  string model  (right).
The $\chi$-squared value per degree of freedom using the dark matter model with a NFW profile is
$0.598$ while that using  the string model is $0.7394$.
}
\end{figure}
\clearpage
\begin{figure}[htbp]
\begin{align*}
\includegraphics[width=0.45\textwidth]{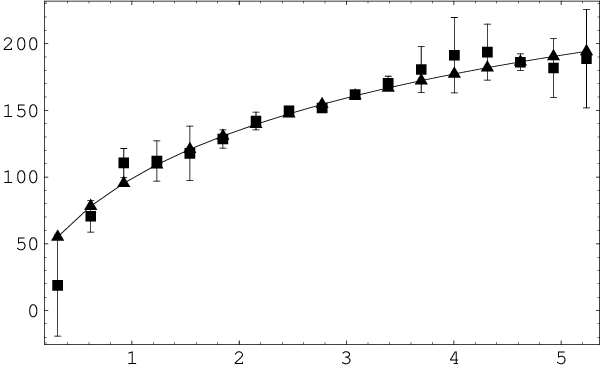}
 &~~~
\includegraphics[width=0.45\textwidth]{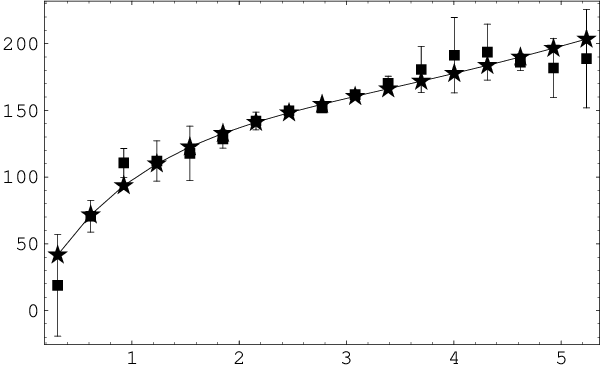}
\end{align*}
\caption{Rotation curve of ngc~4569 fit with the dark matter model   (left) and with the  string model  (right).
The $\chi$-squared value per degree of freedom using the dark matter model with a NFW profile is$0.651$ while that using  the string model is $0.6621$.}
\label{fig:ngc4569}
\end{figure}

\begin{figure}[htbp]
\begin{align*}
 \includegraphics[width=0.45\textwidth]{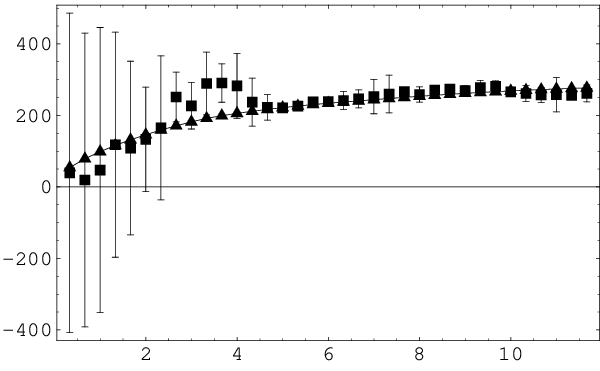}
&~~~
  \includegraphics[width=0.45\textwidth]{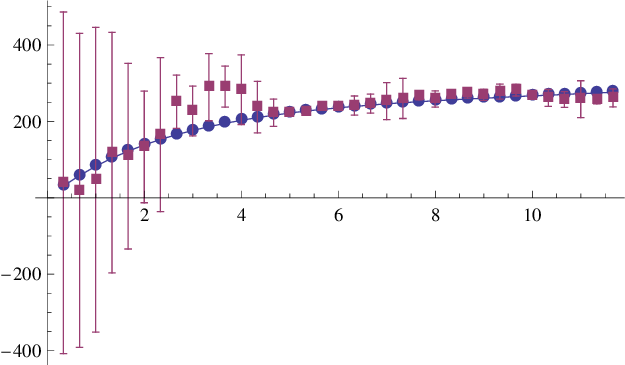}
\end{align*}
\caption{
Rotation curve of NGC~4579 fit with the dark matter model   (left) and with the  string model (right).
The $\chi$-squared value per degree of freedom using the dark matter model with a NFW profile is
$0.719$ while that using  the string model is $0.688$.
}
\label{fig:string_ngc4579}
\end{figure}

\begin{figure}[htbp]
\begin{align*}
\includegraphics[width=0.48\textwidth]{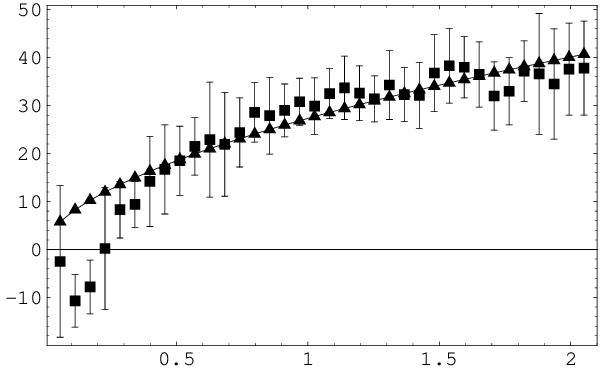}
 &
\includegraphics[width=0.48\textwidth]{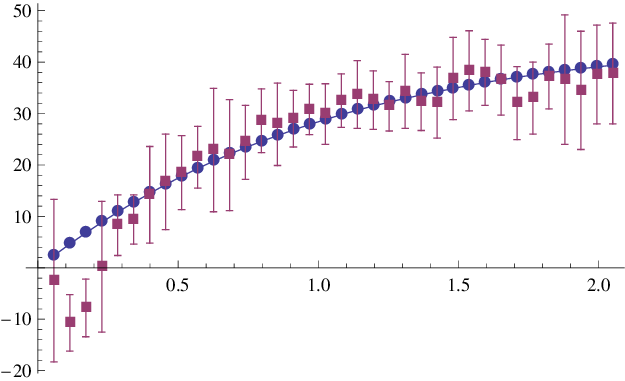}
\end{align*}
\caption{
Rotation curve of ngc~4625 fit with the dark matter model   (left) and with the  string model
(right).
The $\chi$-squared value per degree of freedom using the dark matter model with a NFW profile is
$0.874$ while that using  the string model is $0.5478$.
}
\end{figure}

\begin{figure}[htbp]
\begin{align*}
\includegraphics[width=0.48\textwidth]{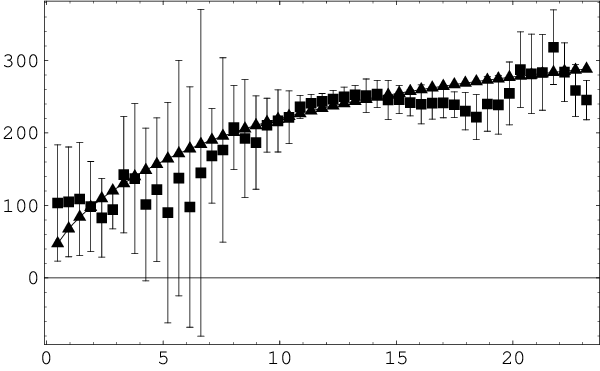}
 &
\includegraphics[width=0.48\textwidth]{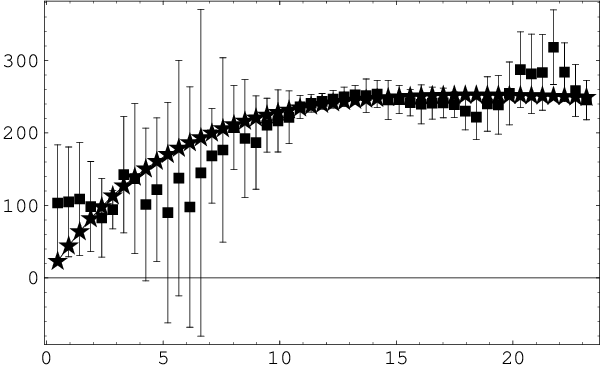}
\end{align*}
\caption{Rotation curve of ngc~4725 fit with the dark matter model   (left) and with the  string  model  (right).
The $\chi$-squared value per degree of freedom using the dark matter model with a NFW profile is
$0.481$ while that using  the string model is $0.2683$.
}
\end{figure}

\begin{figure}[htbp]
\begin{align*}
\includegraphics[width=0.48\textwidth]{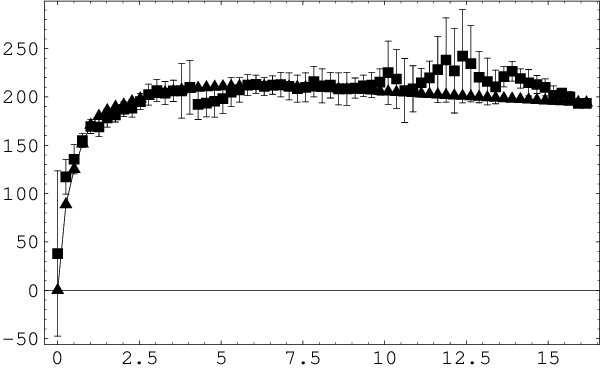}
&
\includegraphics[width=0.48\textwidth]{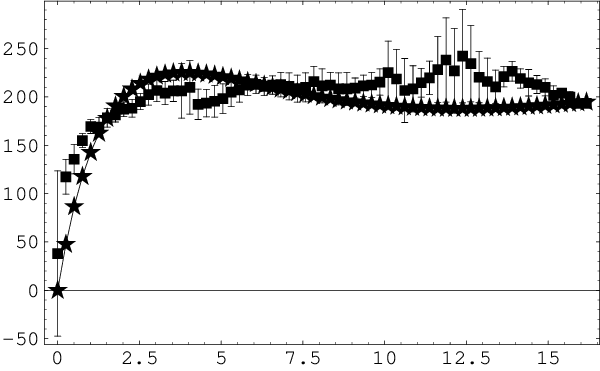}
\end{align*}
\caption{Rotation curve of NGC~5055 fit with the dark matter model   (left) and with the  string model  (right).
The $\chi$-squared value per degree of freedom using the dark matter model with a NFW profile is $1.133$ while that using  the string model is $3.5208$.}
\end{figure}

\begin{figure}[htbp]
\begin{align*}
\includegraphics[width=0.48\textwidth]{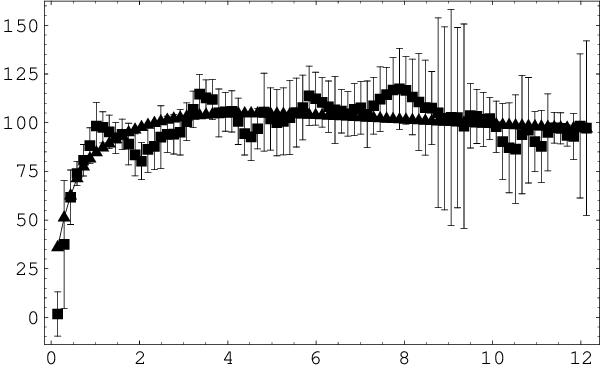}
 &
\includegraphics[width=0.48\textwidth]{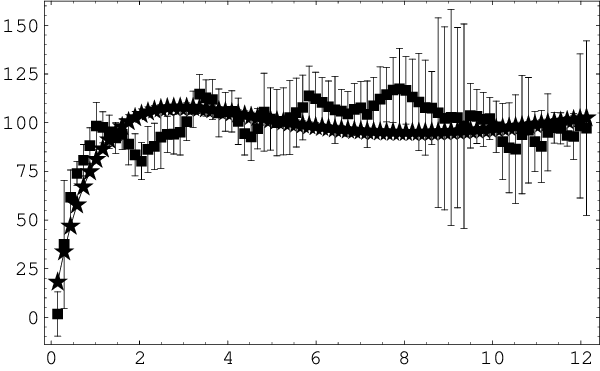}
\end{align*}
\caption{Rotation curve of ngc~5194 fit with the dark matter model   (left) and with the  string
model  (right).
The $\chi$-squared value per degree of freedom using the dark matter model with a NFW profile is
$0.442$ while that using  the string model is $0.8067$.
}
\end{figure}

\begin{figure}[htbp]
\begin{align*}
\includegraphics[width=0.48\textwidth]{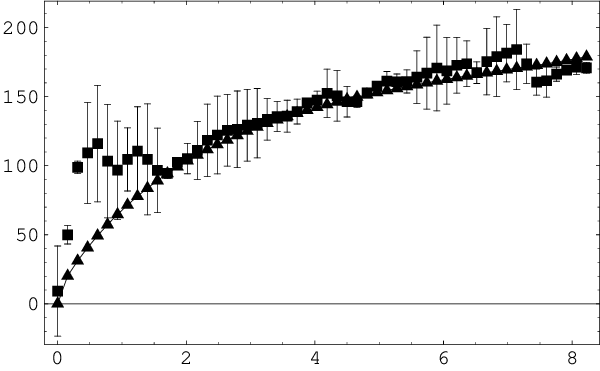} &
\includegraphics[width=0.48\textwidth]{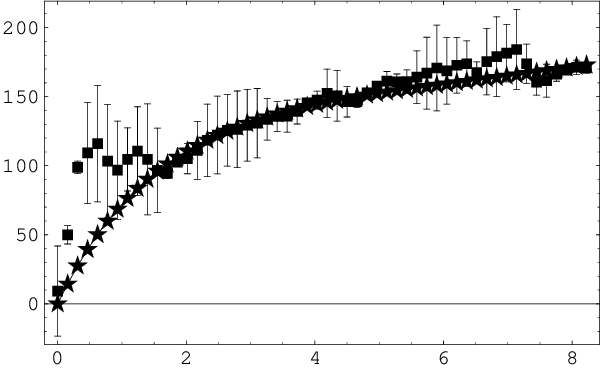}
\end{align*}
\caption{Rotation curve of ngc~6946 fit with the dark matter model   (left) and with the  string  model  (right).
The $\chi$-squared value per degree of freedom using the dark matter model with a NFW profile is
$5.670$ while that using  the string model is $6.0346$.
}
\end{figure}
\begin{figure}[htbp]
\begin{align*}
\includegraphics[width=0.48\textwidth]{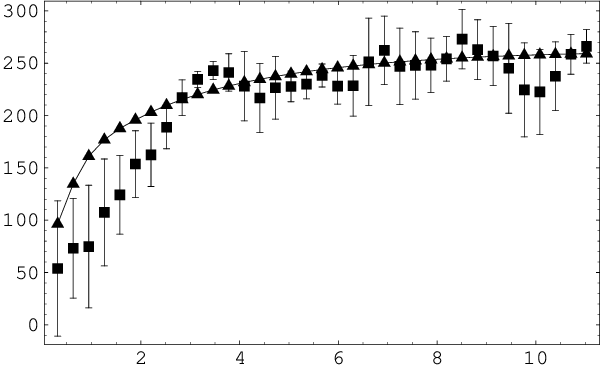}
 &
\includegraphics[width=0.48\textwidth]{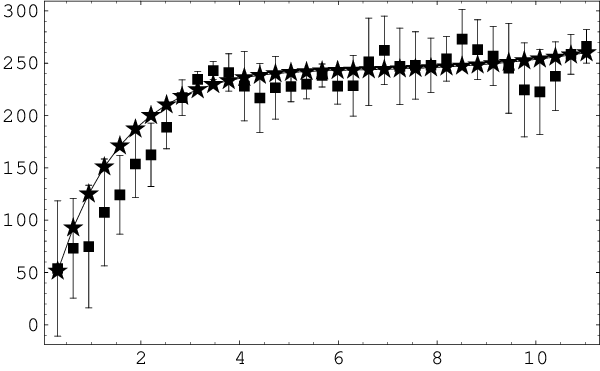}
\end{align*}
\caption{Rotation curve of NGC~7331 fit with the dark matter model   (left) and with the  string  model  (right).
The $\chi$-squared value per degree of freedom using the dark matter model with a NFW profile is
$0.839$ while that using  the string model is $0.5221$.
}
\end{figure}



\clearpage



\end{document}